\RequirePackage{fix-cm}
\documentclass[twocolumn]{svjour3}          
\smartqed  
\usepackage{graphicx}

\usepackage{microtype}                 
\PassOptionsToPackage{warn}{textcomp}  
\usepackage{textcomp}                  
\usepackage{mathptmx}                  
\usepackage{times}                     
\usepackage{cite}                      
\usepackage{tabu}                      
\usepackage{booktabs}                  
\usepackage{multirow}
\usepackage{subfig}
\usepackage{longtable}
\usepackage{xcolor}
\usepackage{hyperref}

\newcommand{\hf}{\hspace*{\fill}}

\newlength{\wl}
\newlength{\wt}
\newlength{\wc}
\newlength{\wta}
\newlength{\wtb}
\newlength{\wtc}
\newlength{\wtd}
\newlength{\wte}
\newlength{\wtf}

\newcommand{\STAB}[1]{\begin{tabular}{@{}c@{}}#1\end{tabular}}
\begin{document}

\title{
On the impact of VR assessment on the Quality of Experience of Highly Realistic Digital Humans
\thanks{This work is funded by the European Commission H2020 program, under the grant agreement 762111, \textit{VRTogether}, http://vrtogether.eu/. On behalf of all authors, the corresponding author states that there is no conflict of interest.}
}
\subtitle{A Volumetric Video Case Study}


\author{Irene Viola        \and
        Shishir Subramanyam \and
        Jie Li \and
        Pablo Cesar
}


\institute{Irene Viola and Jie Li \at
              CWI, Amsterdam, The Netherlands \\
              \email{irene@cwi.nl, jie.li@cwi.nl} \\
              Shishir Subramanyam and Pablo Cesar \at
              CWI, Amsterdam, and TU Delft, Delft, The Netherlands \\
              \email{subraman@cwi.nl, p.s.cesar@cwi.nl}
}

\date{Received: date / Accepted: date}

\maketitle

\begin{abstract}
\textcolor{black}{Fuelled by the increase in popularity of virtual and augmented reality applications, point clouds have emerged as a popular 3D format for acquisition and rendering of digital humans, thanks to their versatility and real-time capabilities. Due to technological constraints and real-time rendering limitations, however, the visual quality of dynamic point cloud contents is seldom evaluated using virtual and augmented reality devices, instead relying on prerecorded videos displayed on conventional 2D screens. In this study, we evaluate how the visual quality of point clouds representing digital humans is affected by compression distortions. In particular, we compare three different viewing conditions based on the degrees of freedom that are granted to the viewer: passive viewing (2DTV), head rotation (3DoF), and rotation and translation (6DoF), to understand how interacting in the virtual space affects the perception of quality. We provide both quantitative and qualitative results of our evaluation involving 78 participants, and we make the data publicly available. To the best of our knowledge, this is the first study evaluating the quality of dynamic point clouds in virtual reality, and comparing it to traditional viewing settings. Results highlight the dependency of visual quality on the content under test, and limitations in the way current data sets are used to evaluate compression solutions. Moreover, influencing factors in quality evaluation in VR, and shortcomings in how point cloud encoding solutions handle visually-lossless compression, are discussed.}
%
\end{abstract}

\section{Introduction}
\label{intro}

\textcolor{black}{Recent technological advances in devices for capturing and rendering immersive media contents, together with the fast processing capabilities of commodity hardware, have fostered the development of new applications for Virtual Reality (VR), Augmented Reality (AR) and Mixed Reality (XR). Such applications usher a new way to engage and interact with media contents: whereas in traditional 2D consumption, users are passive receivers and have limited possibilities of manipulating the contents they are visualizing, immersive media allow for more interactivity in deciding which content should be displayed by each user. Commonly, immersive media applications can be classified based on the Degrees of Freedom (DoF) that are available to the end user to explore the virtual world: 3DoF refers to the availability of only head rotation as a tool for interaction, as for example in omnidirectional imagining, whereas 6DoF refers to the ability to operate translational movements as well as rotational movements in the 3D space.}


\textcolor{black}{In order to populate immersive VR, AR and XR applications, volumetric contents are needed. In this context, point clouds have emerged as a popular format to capture and represent volumetric reconstructions of real-world objects and people, due to their simplicity and versatility. Geometrical representation in point clouds is obtained through a collection of points with $x, y$ and $z$ coordinates in Euclidean space; in addition, attributes such as colour may be included included at each point location. This enables a simple representation that requires no additional pre-processing, is resilient to noise introduced during capture, and enforces no restrictions on the attributes that can be encoded at each point location. However, one main drawback for the deployment of this type of content is the large amount of data that is required in order to produce a photorealistic representation: uncompressed, a single point cloud frame containing one million points requires roughly 20MB to be transmitted. Compression becomes therefore essential for efficient storage and feasible transmission over bandwidth-limited networks. Thus, significant research \cite{mekuria2017design} and industrial \cite{schwarz2018emerging} effort has been focused on optimizing encoding and transmission, as demonstrated by the ongoing standardization endeavors by bodies such as JPEG~\cite{ebrahimi2016jpeg, da2019point} and MPEG~\cite{jang2019video}.}

Given the significant storage and bandwidth requirements for dense dynamic point clouds, decisions need to be taken regarding the delivery (type of encoder, bit-rate) to ensure an acceptable quality of experience, depending on the viewing conditions. \textcolor{black}{In a previous paper~\cite{subramanyam2020comparing}, we analysed the impact of different viewing conditions in VR environments, namely, with 3DoF or 6DoF locomotion. With this work, we aim to extend our previous analysis by including results and discussions obtained in a baseline viewing condition using traditional 2D screens (2DTV), which is by large the most commonly used environment for user studies for point cloud quality assessment.} 

\textcolor{black}{In this paper, we report findings obtained in a user study involving 78 participants assessing 72 stimuli based on eight dynamic point clouds sequences depicting humans. Each point cloud sequence was compressed using two encoding solutions at 4 bit-rates, and evaluated in three viewing conditions (6DoF, 3DoF and 2DTV). The gathered data include rating scores, presence questionnaires, simulator sickness reports, along with average watching time.}
Contributions of the paper are three-fold: 
\begin{enumerate}
        \item An extensive evaluation of the quality of highly realistic digital humans represented as dynamic point clouds in immersive and traditional TV viewing conditions is provided. Existing protocols \cite{alexiou2018impact, alexiou2017towards, Alexiou2019comprehensive, zhang2014subjective, zerman2019subjective} did not consider the dynamic nature of the point clouds, focused on one type of data set, did not take into account VR viewing conditions, \textcolor{black}{and did not compare VR findings with 2DTV conditions using dynamic contents;} 
        \item Quantitative subjective results about the perceived quality of the contents, along with qualitative insights on what is important for users in interacting with digital humans in VR, are presented. Such results will help in better configuring the network conditions for the delivery of points clouds for real-time transmission, and have implications over ongoing research and standardisation work regarding the underlying compression technology;
        \textcolor{black}{\item The collected raw data, which is comprised of rating scores, presence questionnaires, and simulator sickness reports, is made available to the research community, along with scripts to faithfully recreate the stimuli under exam\footnote{\url{https://github.com/cwi-dis/2DTV_VR_QoE}}. This will aid reproducibility, while contributing to ongoing research in the area.}
\end{enumerate}

\textcolor{black}{The remainder of the paper is organised as follows. Section~\ref{sec:sota} summarises the related work in the field of point cloud compression and subjective visual quality assessment. Section~\ref{sec:method} details the methodology that was followed to conduct the experiments and analyse the data. In Section~\ref{sec:results}, we report the quantitative and qualitative results of the subjective visual quality assessment, along with commenting the findings in terms of simulator sickness, presence, and interaction time. Key factors and issues for visual quality assessment of dynamic point clouds are discussed in Section~\ref{sec:discussion}. Finally, Section~\ref{sec:conclusions} concludes the paper. Additional data regarding the statistical analysis of the results is offered in Appendix~\ref{sec:appendix}.}


\begin{figure*}[!ht]
		\centering
		\hf
		\subfloat[\emph{V-PCC}]{\includegraphics[width=0.7\columnwidth]{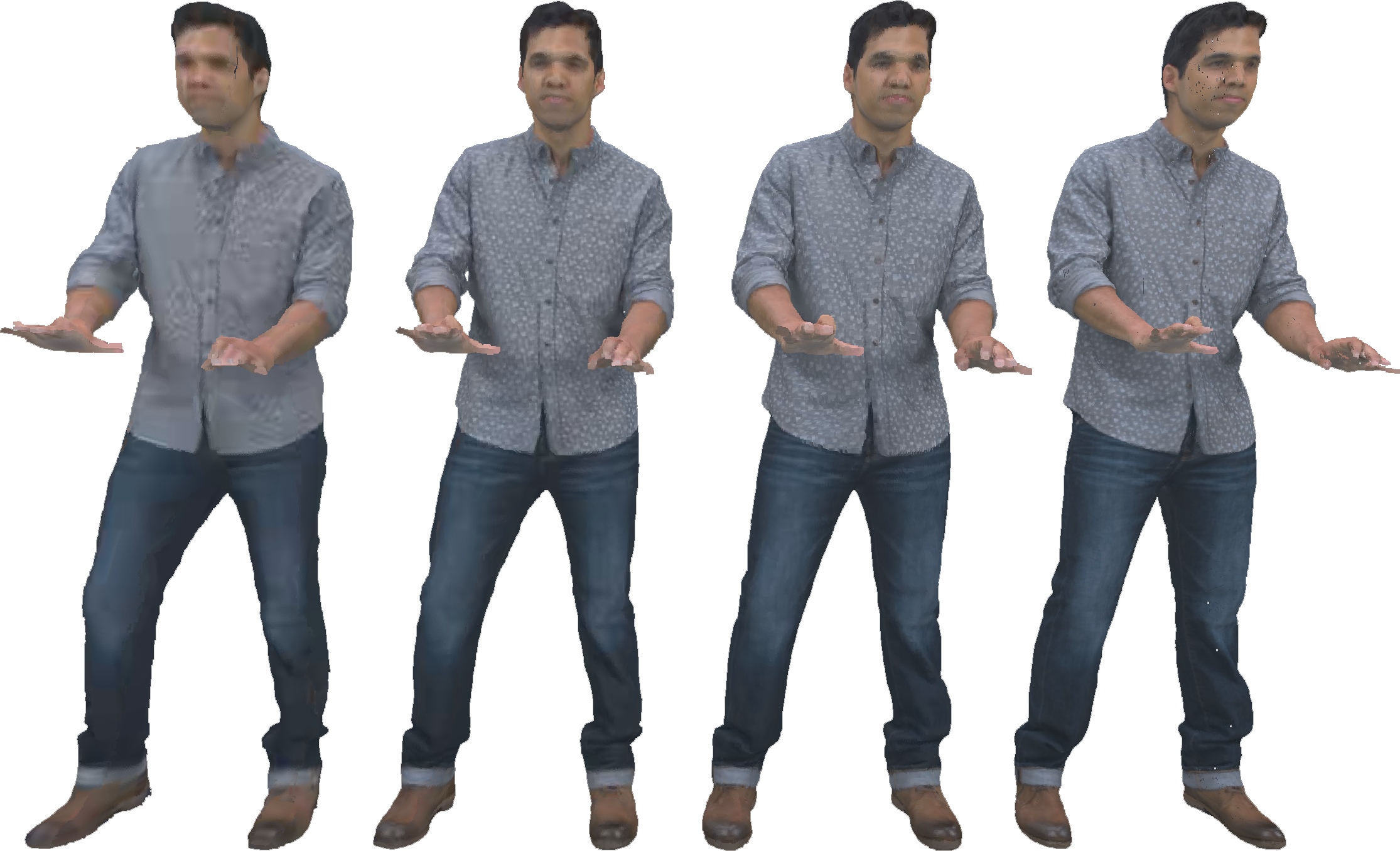}}\hf \hf
\subfloat[\emph{MPEG Anchor}]{\includegraphics[width=0.75\columnwidth]{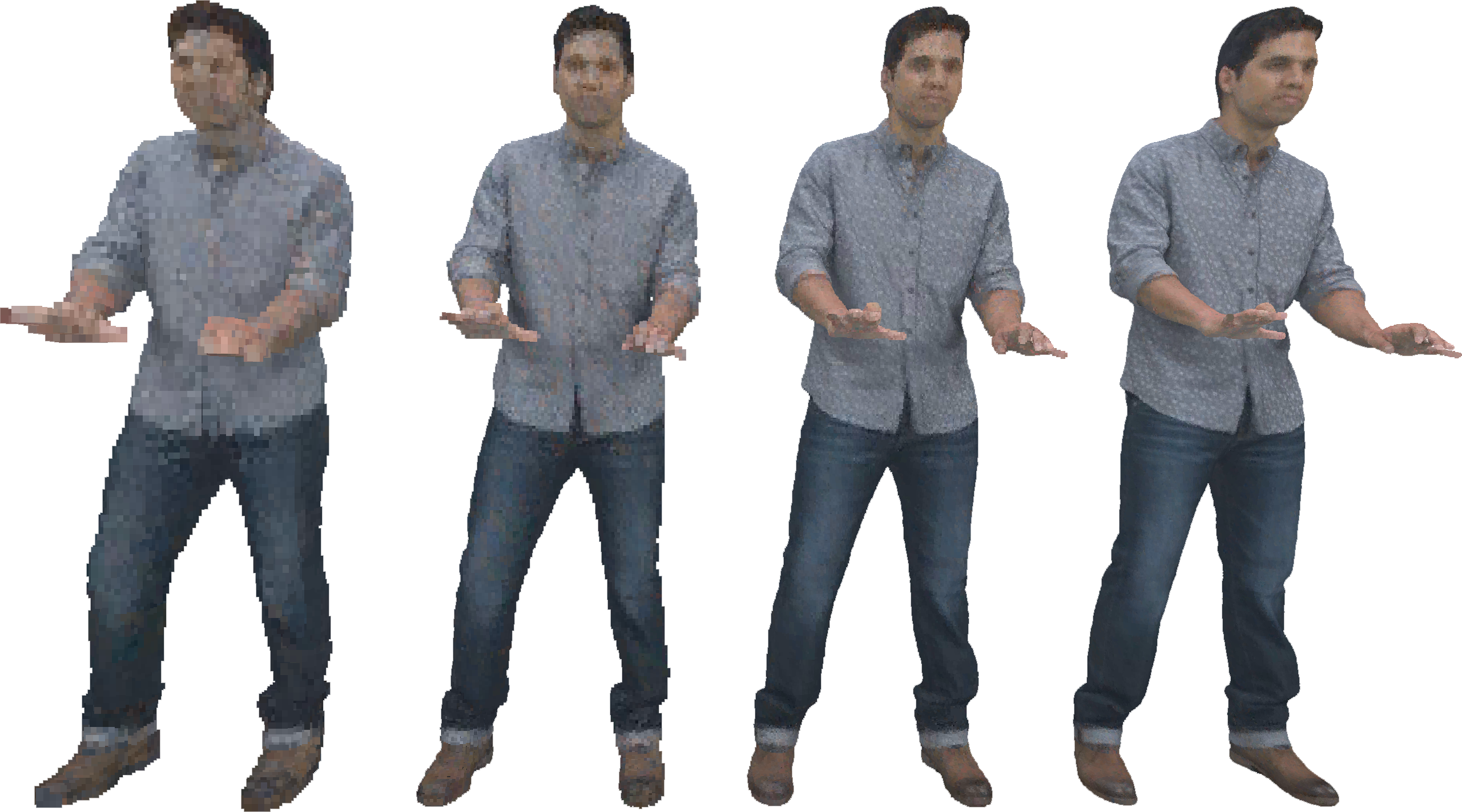}}\hf
		\caption{Point Cloud Digital Humans compressed using two point cloud codecs, V-PCC (left) and MPEG anchor (right), at the 4 selected bit-rates.}
		\label{fig:pcccreation}
		\vspace{-0.5cm}
\end{figure*}

\section{Related work}
\label{sec:sota}

\subsection{Quality assessment for point clouds}

There is a growing interest on subjective quality assessment of point clouds rendered on 2D displays. Zhang et al. \cite{zhang2014subjective} evaluated the quality degradation effect of resolution, shape and color on static point clouds. The results indicate that resolution is almost linearly correlated with the perceived quality, and color has less impact than shape on the perceived quality. Zerman et al. \cite{zerman2019subjective} compressed two dynamic human point clouds using a state-of-the-art algorithm 
\cite{mammou2017pcc}, and assessed the effects of this algorithm and input point counts on the perceived quality. Their results showed that no direct correlation was found between human viewers' quality ratings and input point counts. In a recent study \cite{da2019point}, a protocol to conduct subjective quality evaluations and benchmark objective quality metrics were proposed. The viewers passively assessed the quality of a set of static point clouds, as animations with pre-defined movement path. 
In a comprehensive work by Alexiou et al.\cite{Alexiou2019comprehensive}, the entire set of emerging point cloud compression encoders developed in the MPEG committee were evaluated through a series of subjective quality assessment experiments. 
Nine static models, including both humans and objects, were used in the experiments. The experiments provided insights regarding the performance of the encoders and the types of degradation they introduce. 
Zerman et al.~\cite{zerman2020textured} compared the visual quality of point cloud and mesh contents compressed using state-of-the-art algorithms, concluding that, while meshes are more suitable for high bitrate streaming, point cloud compression appears to be more advantageous at lower bitrates. Perry et al.~\cite{perry2020quality} conducted an experiment in 4 laboratories, comparing the latest standardized compression solutions on static point cloud contents. 

Only a limited number of point cloud quality assessment studies have been conducted in immersive environments. Mekuria et al. \cite{mekuria2017design} evaluated the subjective quality of their codec performance in a realistic 3D tele-immersive system, in which users were represented as 3D avatars and/or 3D dynamic point clouds, and could navigate in the virtual space using mouse cursor in a desktop setting. Several aspects of quality, such as level of immersiveness, togetherness, realism, quality of motion, were considered. 
Alexiou and Ebrahimi \cite{alexiou2017towards} proposed the use of AR to subjectively evaluate the quality of colorless point cloud geometry. Zerman et al.~\cite{zerman2021user} presented a behaviour analysis of users interacting with colored volumetric media in a AR application.
Tran et al. \cite{tran2019subjective} suggested that, in case of evaluating video quality in an immersive setup, aspects such as cybersickness and presence should not be overlooked. 
Recently, an evaluation of static point cloud contents was conducted in a VR environment~\cite{alexiou2020pointxr}. 

\textcolor{black}{Our work aims at comparing different viewing paradigms for dynamic point clouds, namely 6DoF, 3DoF, and 2DTV conditions. Such a comparison is largely absent in the literature, as previous work has mainly focused on static point cloud contents and single viewing conditions.}




\subsection{Point cloud compression}
A single point cloud frame is represented by an unordered collection of points sampled from the surface of an object. In a dynamic sequence of point clouds, there are no correspondences of points maintained across frames. Thus, detecting spatial and temporal redundancies is often difficult, making point cloud compression challenging. Octrees have been used extensively as a space partitioning structure to represent point cloud geometry~\cite{meagher1982geometric, jackins1980oct}. They are a 3D extension of the 2D quadtree used to encode video and images.

Research into point cloud compression can be broadly divided into two categories: model-based and projection-based. The first uses signal processing or deep learning techniques to compress either the geometrical composition of the point cloud, or its attributes, such as color. Zhang et al. \cite{zhang:gft} proposed a method to compress point cloud attributes using a Graph Fourier Transform. They assume that an octree has been created and separately coded for geometry prior to coding attributes. De Queiroz and Chou \cite{queiroz:raht} used a Region Adaptive Hierarchical Transform (RAHT) to use the colors of nodes in lower levels of the octree to predict the colors of nodes in the next level. In~\cite{cohen2016point}, authors adopt techniques from traditional image and video processing, using 3D block prediction in combination with shape-adaptive DCT and graph transforms.

The second category of point cloud codecs aim at projecting the point cloud information onto a 2D canvas, subsequently using legacy image and video compression solutions to encode them. Intra Frame coding in octrees can be achieved by entropy coding the occupancy codes, and then compress the color attributes by mapping them to a 2D grid and using legacy JPEG image compression, as shown in \cite{mekuria2017design}.
In 2017, MPEG started a standardization activity to determine a new standard codec for point clouds. They used the codec created by Mekuria et al. \cite{mekuria2017design} as an anchor to evaluate proposals. \textcolor{black}{To encode dynamic point cloud sequences, MPEG has currently standardized one method for dynamic dense point clouds, namely V-PCC~\cite{MPEG-VPCC-standard_ISO}, and is in the process of standardizing one method for dynamically acquired, sparse point clouds, namely G-PCC~\cite{MPEG-GPCC-standard_ISO}.} 

\textcolor{black}{Recently, deep learning solutions for point cloud compressions have been proposed, to encode either geometry or color information, or a combination thereof. Quach et al.~\cite{quach2019learning} propose the use of an auto-encoder to efficiently compress geometry information, and they subsequently analyse the impact of several parameters on the performance~\cite{quach2020improved}. Similarly, Guarda et al.~\cite{guarda2019point} propose a Convolutional Neural Network (CNN) architecture to encode and decode point cloud contents. They further analyse its performance in~\cite{guarda2019deep}, and extend the work in~\cite{guarda2020deep} by employing explicit and implicit quantization. A deeper architecture is proposed in~\cite{wang2021lossy}, which uses 3D convolutional layers along with variational auto-encoders to achieve favorable compression efficiency. In~\cite{alexiou2020towards}, Alexiou et al. propose a deep learning architecture to encode both geometry and color attributes, and analyse the performance of various parameters on the coding efficiency and visual quality.}

\textcolor{black}{A complete survey of point cloud compression solutions can be found here~\cite{cao2021compression}. In our work, we elected to adopt the MPEG Anchor that was used to evaluate the Call for Proposals for the MPEG standardization efforts in point cloud compression~\cite{schwarz2018emerging}, and the MPEG standard for dynamic point clouds V-PCC~\cite{MPEG-VPCC-standard_ISO}, as they have both been widely used in quality evaluation campaigns in the literature. }


\section{Methodology}
\label{sec:method}
\subsection{Dataset Preparation}
\begin{figure*}[t]
		\centering
		\includegraphics[width = 0.7\textwidth]{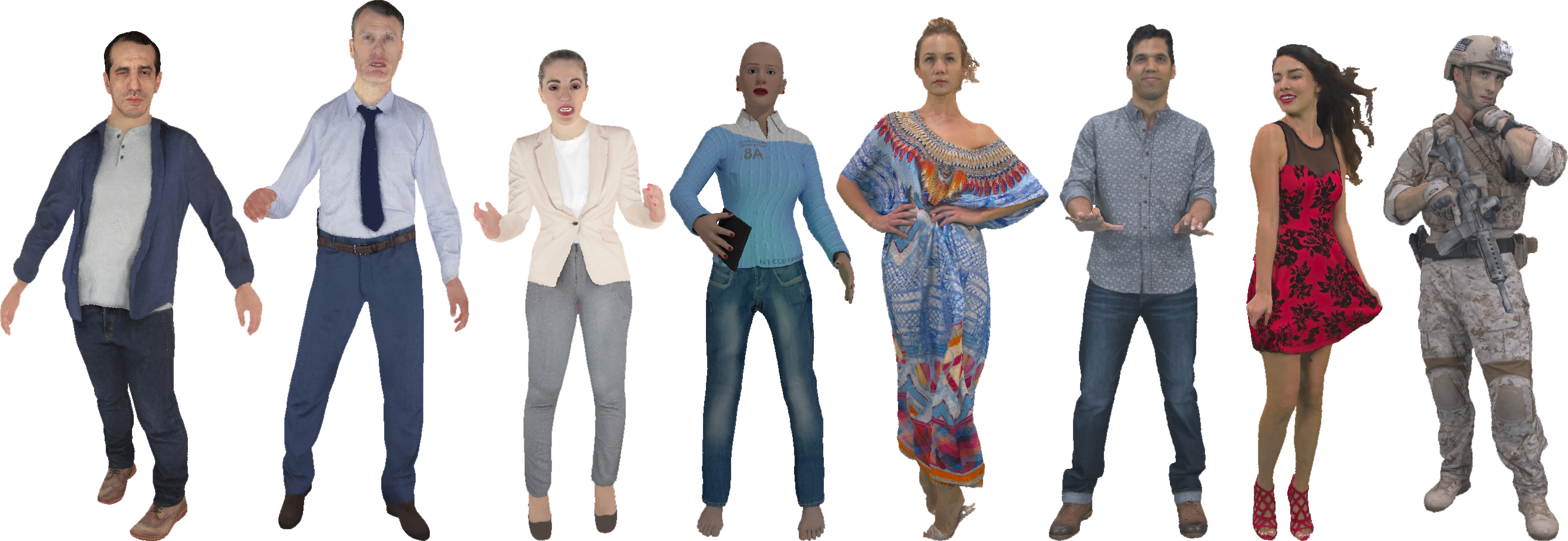}
		\caption{{Sequences used for the test, from left to right: Manfred, Sarge, Despoina, Queen, Longdress, Loot, Red and black, Soldier}}
		\label{fig:pcdataset}
		\vspace{-0.5cm}
	\end{figure*}

A dataset of dynamic point cloud sequences was used from the MPEG repository. All sequences were clipped to five seconds and sampled at 30 frames per second. {This included point cloud sequences~\cite{8i:dataset} captured using photogrammetry (\textit{Longdress, Loot, Red and black, Soldier}, shown in figure \ref{fig:pcdataset}) and one sequence of a synthetic character sampled from an animated mesh (\textit{Queen}).
Four additional point cloud sequences; \textit{Manfred, Despoina, Sarge} (shown in Figure \ref{fig:pcdataset}) were added for the evaluation. These sequences were created using motion-captured animated mesh sequences. }
	
Keyframes were selected at 30 frames per second and extracted along with the associated mesh materials. Particular care was put in ensuring the selected sequences have the characters facing the user and speaking in their general direction. Then, 1 million points were randomly sampled, independently per key frame to create a consistent groundtruth dataset. The points were sampled from the mesh surface with a probability proportional to the area of the underlying mesh
face. This was done to ensure no direct point correspondences across point cloud frames, to mimic realistic acquisition and maintain consistency with the rest of the dataset. 
The X, Y, Z coordinates of each point was represented using an unsigned integer, as is required for the current version of the V-PCC software.  Texture information was encoded as 8-bit RGB.

The contents were compressed using two widely available codecs: the MPEG V-PCC codec in the Release 7.0~\cite{schwarz2018emerging} (\textit{C1}), as the state-of-the-art solution in point cloud compression, and the MPEG anchor codec~\cite{mekuria2017design} (\textit{C2}), as a baseline release with real-time capabilities. Bitrate points were selected based on the provided presets for \textit{C1}, to ensure fair use.
For the additional point cloud sequences for which no configuration file was available, the one provided by MPEG for the \textit{Queen} sequence was used for all the contents. We selected the rate points 1, 3 and 5 from the provided preset V-PCC configurations and extended it with an additional final rate point using a texture Quantization Parameter (QP) of 8, a geometry QP of 12, and an occupancy precision of 2. We re-label the rate points as R1, R2, R3 and R4, respectively. All sequences are encoded using the C2AI (Category 2 All Intra) config. For the photogrammetry sequences, we use the predefined dedicated configuration files for each sequence, at the same rate points. 

\textit{C2} is used in an all-intra configuration to match the bit-rates per sequence and rate point (R1-R4) with a tolerance of 10\%, as defined in the MPEG call for proposals. 
{We use an octree depth from 7 to 10 for the rate points R1 to R4, respectively. The highest possible JPEG quantization parameter values were then chosen per sequence, while meeting the target bit rate set using \textit{C1}.}\textcolor{black}{An example of content \textit{Loot} encoded with the two compression solutions at the selected rate points is shown in Figure~\ref{fig:pcccreation}.}

\subsection{Experiment setup}
All point cloud sequences were rendered using the Unity game engine, by storing all the points of each frame in a vertex buffer, and then drawing procedural geometry on the GPU. {The point clouds were rendered using a quadrilateral at each point location with a fixed offset of 0.08 units (this corresponds to a side length of approximately 2mm) around each point (placed at the centre) for all the sequences, to be consistent. }In the case of bitrate R1 generated using the MPEG anchor, we increased the offset value to 0.16 by eye, as the resulting point clouds were too sparse (shown in Figure \ref{fig:pcccreation}b). We maintain a fixed frame rate of 30fps throughout the experiment. 

\textcolor{black}{Three viewing conditions were selected for comparison: 6DoF, 3DoF, and 2DTV condition.} For the 6DoF and 3DoF viewing conditions, participants were asked to wear an Oculus Rift CV1 HMD to view each of the point cloud sequences. For the 3DoF condition, participants were asked to sit on a swivel chair placed at a fixed location in the room and navigate using head movements alone, whereas for the 6DoF condition, participants were allowed to navigate freely within the room. Each sequence was 5 seconds long, after which the playback looped around. 
We set the background of the virtual room to mid-grey, to avoid distractions.
The Oculus Guardian System was used to display in-application wall and floor markers if the participants got too close to the boundary. We used a workstation with 2 GeForce GTX 1080 Ti in SLI for the GPU and an Intel Core i9 Skylake-X 2.9GHz CPU.
\textcolor{black}{For the 2DTV condition, the videos were created offline, using the same rendering as the other two viewing condition, and played back to the users using MPV\footnote{\url{https://mpv.io/}}. A 25" Dell UltraSharp U2515H QHD (2560x1440 px) monitor was used to display the videos. The monitor was calibrated using an i1Display Pro color calibration device according to the following profile: sRGB Gamut, D65 white point, 120cd/m2 brightness, and minimum black level of 0.2 cd/m2. The test was performed in a room with controlled lighting and mid-grey walls, in accordance with ITU-T Recommendation BT.500-13~\cite{ITURBT500}.The illumination level measured on the screens was 15 lux.}

\subsection{Subjective methodology}
To perform the experiments, the subjective methodology Absolute Category Rating with Hidden References (ACR-HR) was selected, according to ITU-T Recommendations P.910~\cite{ITUTP910}. Participants were asked to observe the video sequences depicting digital humans, and rate the corresponding visual quality on a scale from 1 to 5 (\textit{1-Bad}, \textit{2-Poor}, \textit{3-Fair}, \textit{4-Good}, and \textit{5-Excellent}). 

A series of pilot studies were conducted to determine the positioning of digital humans in the virtual space and the length of each sequence, to ensure the sequences were running smoothly within the limited  computer RAM. Due to the huge size of the test material, it was not possible to evaluate all 8 point cloud contents in one single session, as long loading times would have brought fatigue to the participants and corrupted the results. Thus, we decided to split the evaluation into two separate tests: one focused on the evaluation of contents obtained from random sampling of meshes (\textbf{T1}: contents \textit{Queen}, \textit{Manfred}, \textit{Despoina} and \textit{Sarge}), and one focused on contents acquired through photogrammetry (\textbf{T2}: contents \textit{Long dress}, \textit{Soldier}, \textit{Red and black}, and \textit{Loot}). From each sequence, a subset of frames comprising 5 seconds was selected. 

Before the test took place, 3 training sequences depicting examples of \textit{1-Bad}, \textit{5-Excellent} and \textit{3-Fair} were shown to the users to help them familiarize with the viewing condition and test setup, and to guide their rating. Following ITU-T Recommendation BT.500-13~\cite{ITURBT500}, the training sequences were created using one additional content not shown during the test, to prevent biased results. 
Each content sequence was encoded using the point cloud compression algorithms under test.

For each test and viewing condition, 36 stimuli were evaluated. For each stimulus, the 5 second sequence was played at least once in full, and kept in loop until the participants gave their score. The order of the displayed stimuli was randomized per participant and per viewing condition, and the same content was never displayed twice in a row to avoid bias. Moreover, the presentation order of viewing conditions was randomized between participants, to prevent any confounding effect. Two dummy samples were added at the beginning of each viewing session to ease participants into the task, and the corresponding scores were subsequently discarded. 

After each VR viewing conditions (6DoF and 3DoF), participants were requested to fill in the Igroup Presence Questionnaire (IPQ)\cite{schubert2003sense} on a 1-7 discrete scale (1=fully disagree to 7=totally agree) and Simulator Sickness Questionnaire (SSQ) on a 1-4 discrete scale (1=none to 4=severe)\cite{kennedy1993simulator}. IPQ has three subscales, namely Spatial Presence (SP), Involvement (INV) and Experienced Realism (REAL), and one additional general item (G) not belonging to a subscale, which assesses the general "sense of being there", and has high loadings on all three factors, with an especially strong loading on SP \cite{schubert2003sense}. SSQ was developed to measure cybersickness in computer simulation and was derived from a measure of motion sickness \cite{kennedy1993simulator}. For both T1 and T2, after the two viewing conditions, participants were interviewed to 1) compare their experiences of assessing quality in 3DoF and 6DoF, and 2) reflect on the factors they considered when assessing the quality.

For the 6DoF and 3DoF viewing conditions, a total of 27 participants were recruited for T1 (12 males, 15 female,  average age: 22,48 years old), whereas 25 participants were recruited for T2 (17 males, 8 females, average age: 28,39 years old). 
\textcolor{black}{The 2DTV viewing condition was conducted separately, with 26 participants for both T1 and T2 (17 males, 9 females, average age: 34,76 years old).} All participants were screened for color vision and visual acuity, using Isihara and Snellen charts, respectively, according to ITU-T Recommendations P.910~\cite{ITUTP910}.

\subsection{Data analysis}
Outlier detection was performed separately for each of the test datasets T1 and T2, following ITU-T Recommendations P.913~\cite{ITUTP913}. The recommended threshold values of $r_1 = 0.75$ and $r_2 = 0.8$ were used. One outlier was found in test dataset T1, and the corresponding scores were discarded. No outliers were found in the scores collected for T2. Since the 2DTV viewing condition was tested with a different subject population, outlier detection was conducted separately. No outlier was detected for this viewing condition.

After outlier detection, the Mean Opinion Score (MOS) was computed for each stimulus, independently per viewing condition. The associated $95\%$ Confidence Intervals (CIs) were obtained assuming a Student's t-distribution. Additionally, the Differential MOS (DMOS) was obtained by applying HR removal, following the procedure described in ITU-T Recommendations P.913~\cite{ITUTP913}. 
Non-parametric statistical analysis was then used to analyze if there are statistical differences among variables, {using the MATLAB Statistics and Machine Learning Toolbox, along with the ARTool package in R~\cite{matthew_kay_2019_2556415, elkin2021aligned}}. 
\begin{figure*}[!t]
\begin{center}
\hf
\subfloat[\emph{Manfred}]{\includegraphics[width=\wtc]{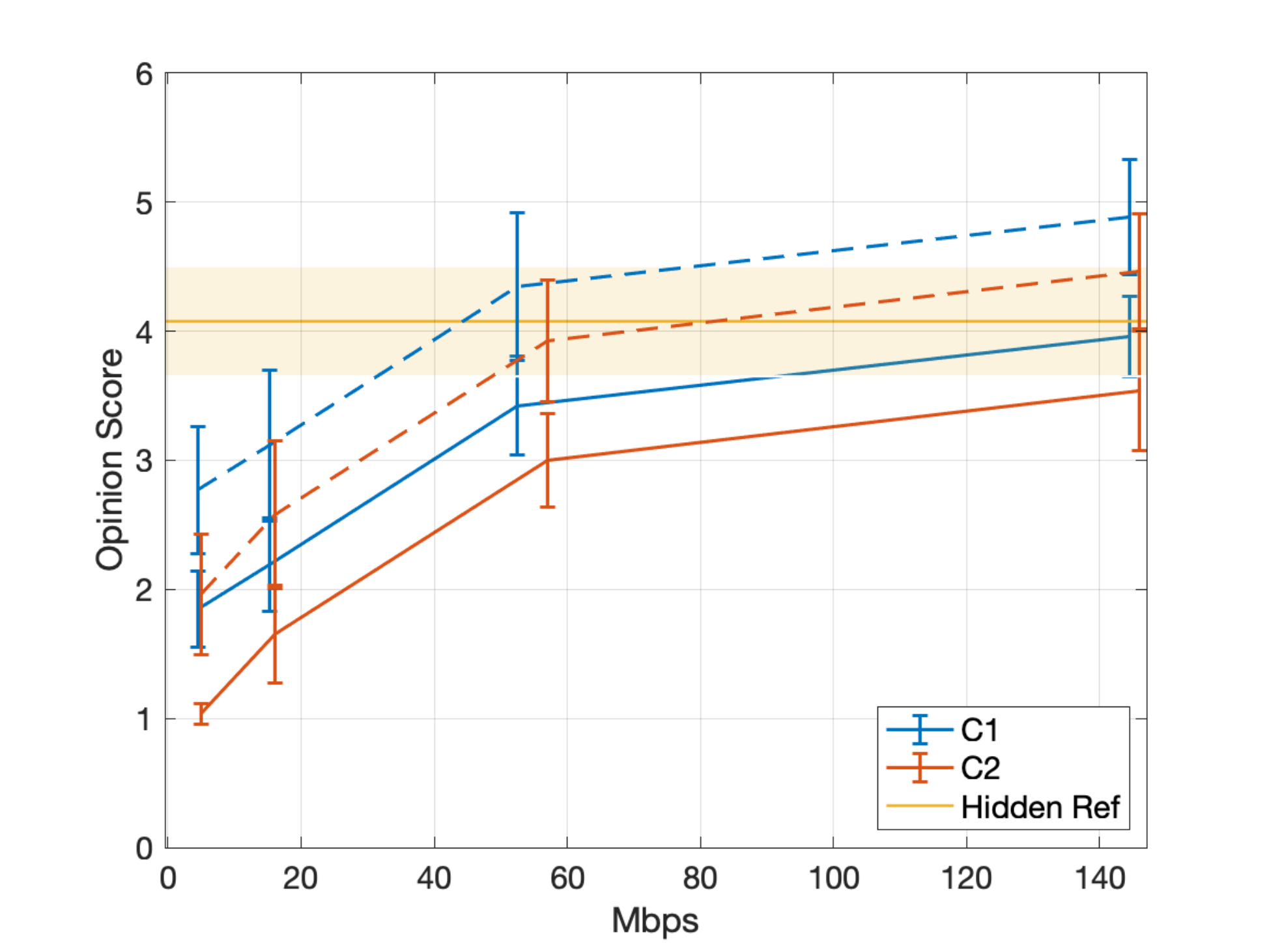}}\hf
\subfloat[\emph{Sarge}]{\includegraphics[width=\wtc]{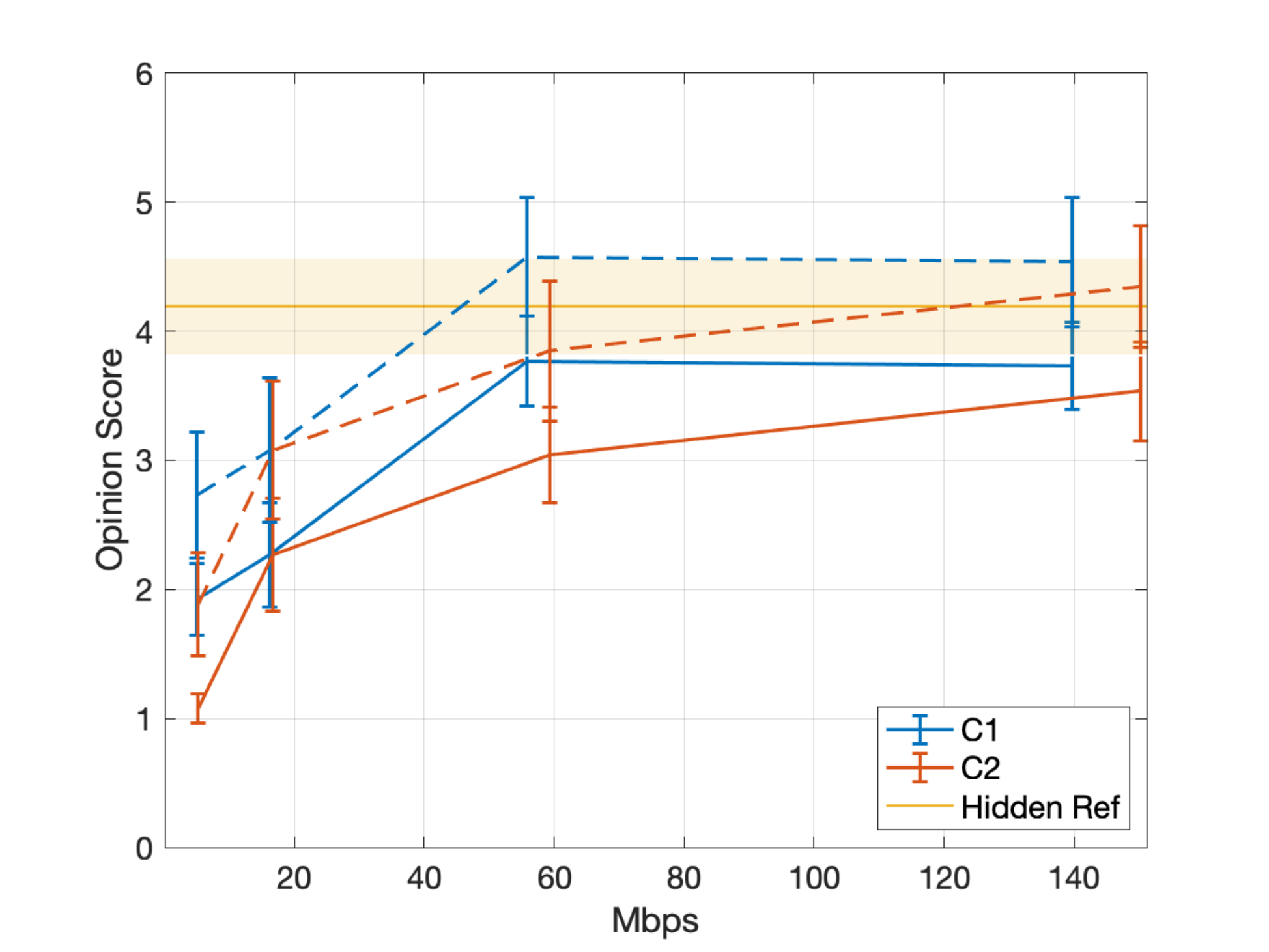}}\hf
\subfloat[\emph{Despoina}]{\includegraphics[width=\wtc]{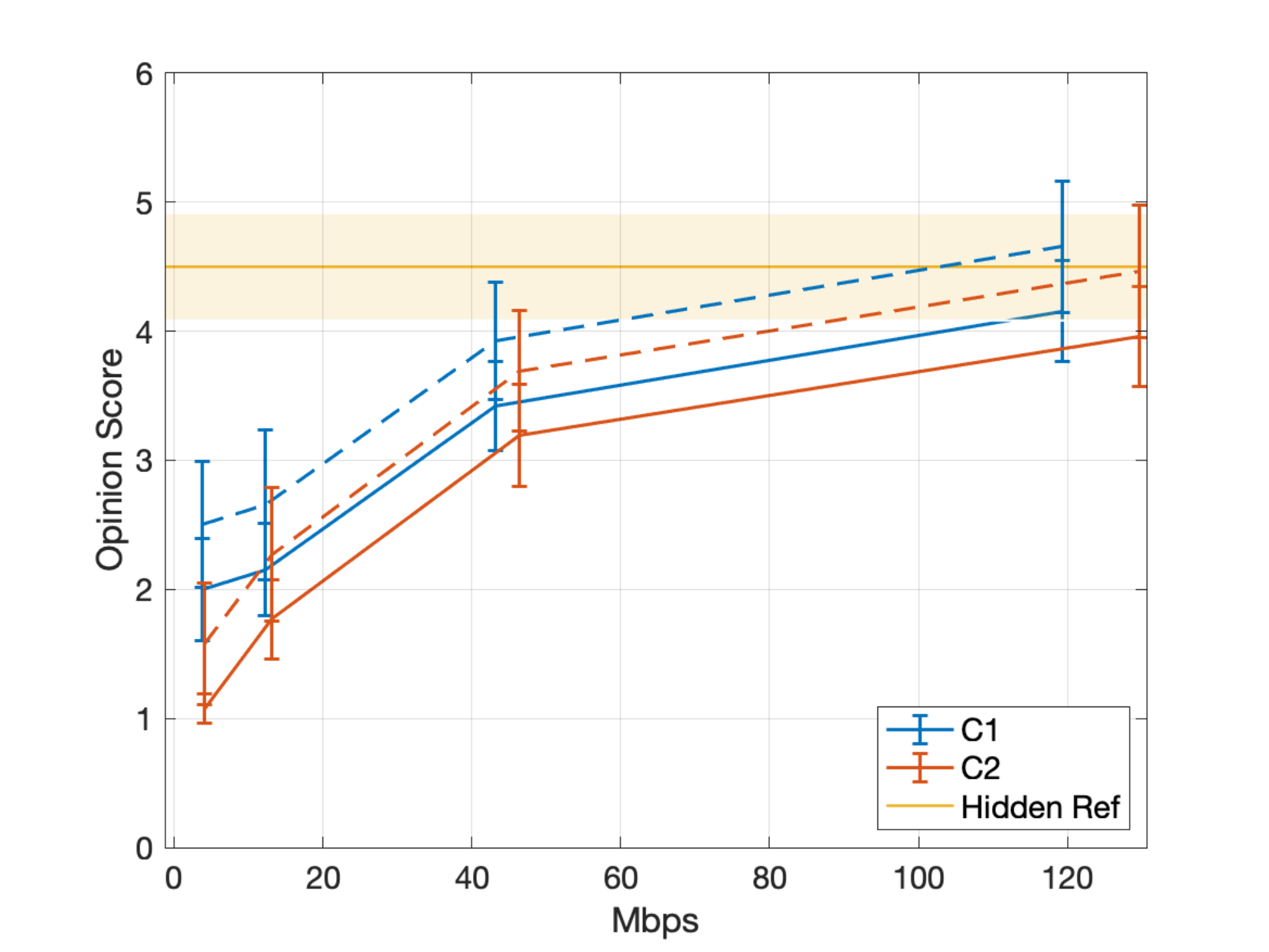}}\hf
\subfloat[\emph{Queen}]{\includegraphics[width=\wtc]{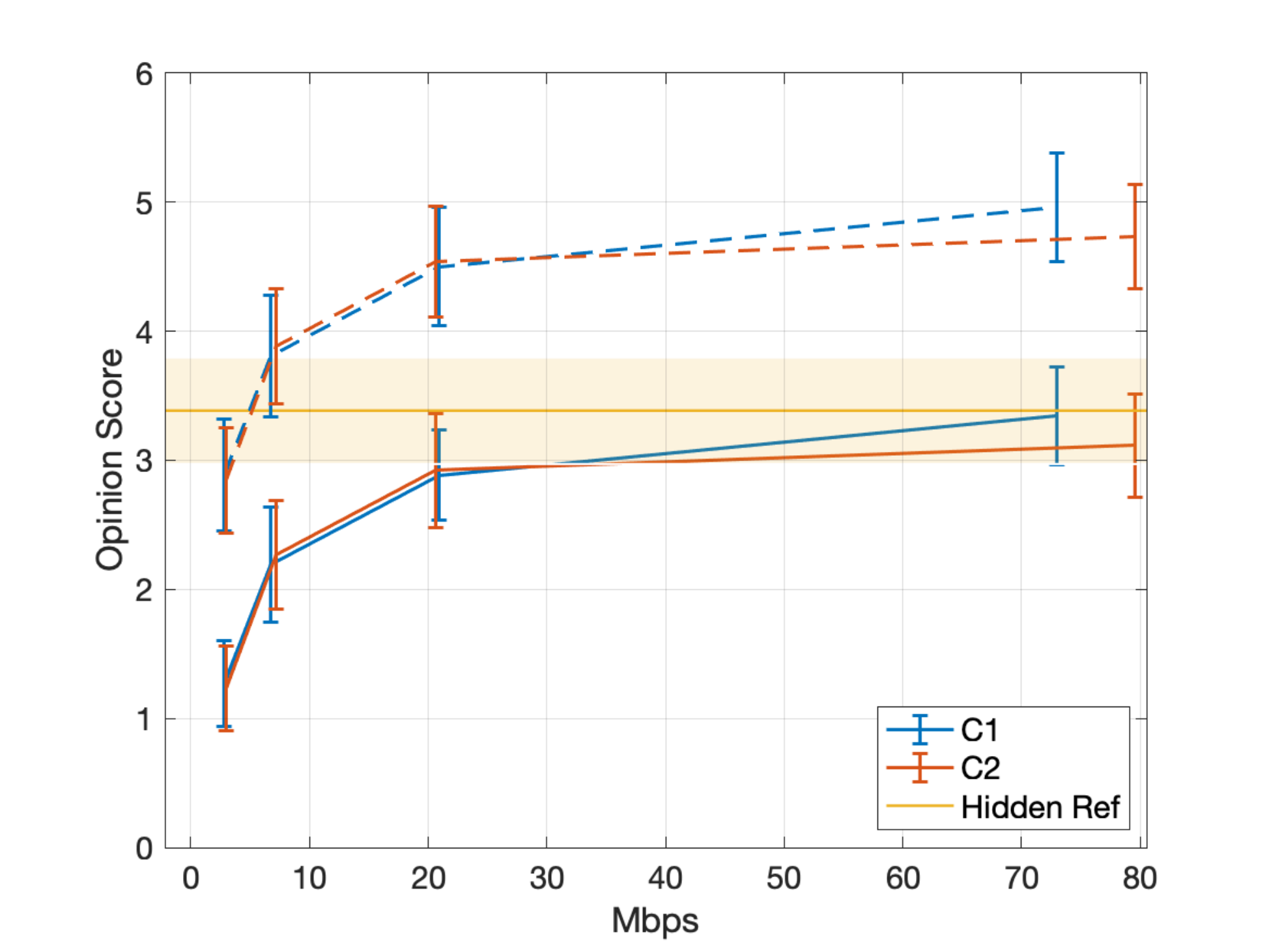}}\hf

\subfloat[\emph{Manfred}]{\includegraphics[width=\wtc]{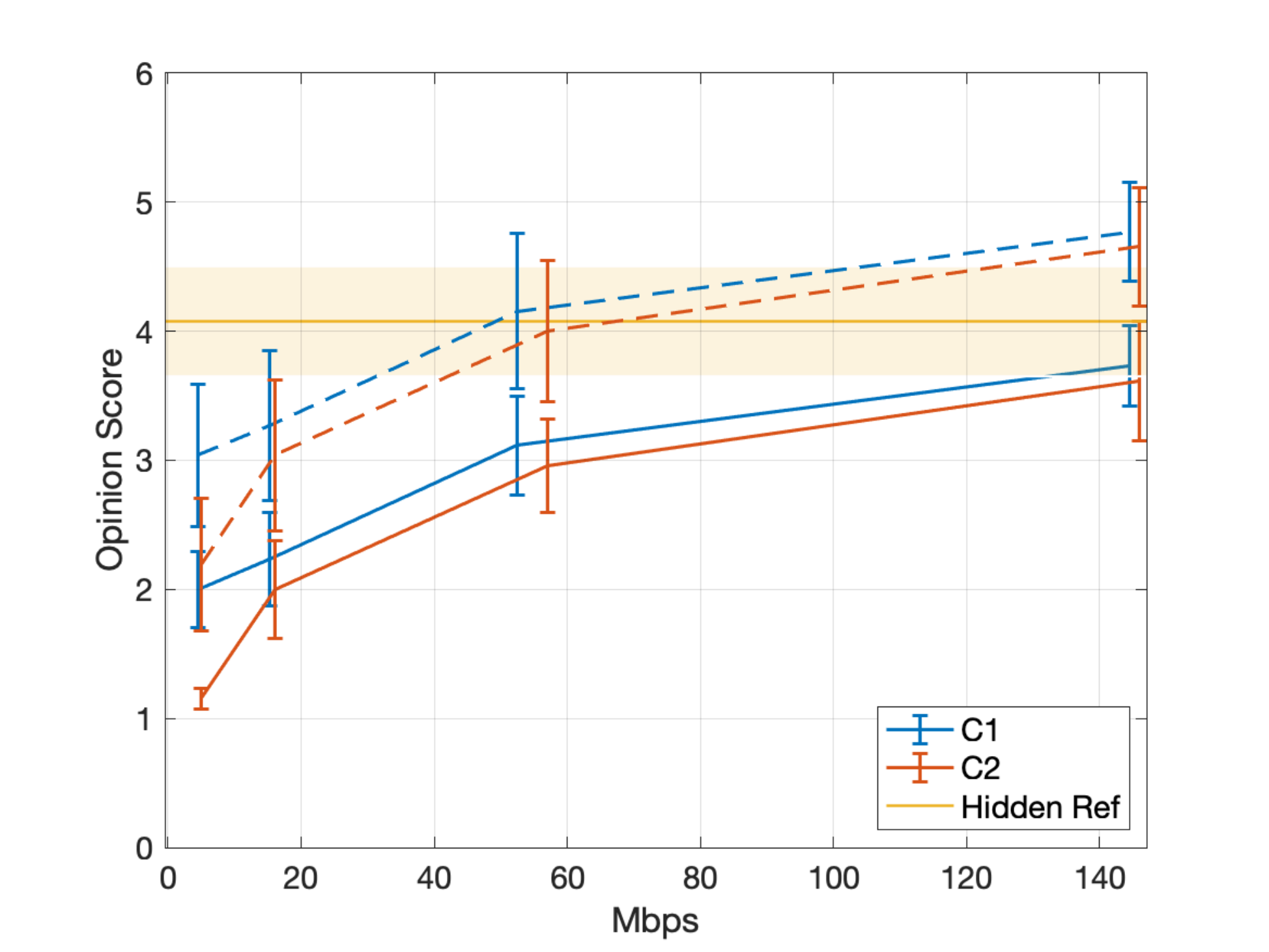}}\hf
\subfloat[\emph{Sarge}]{\includegraphics[width=\wtc]{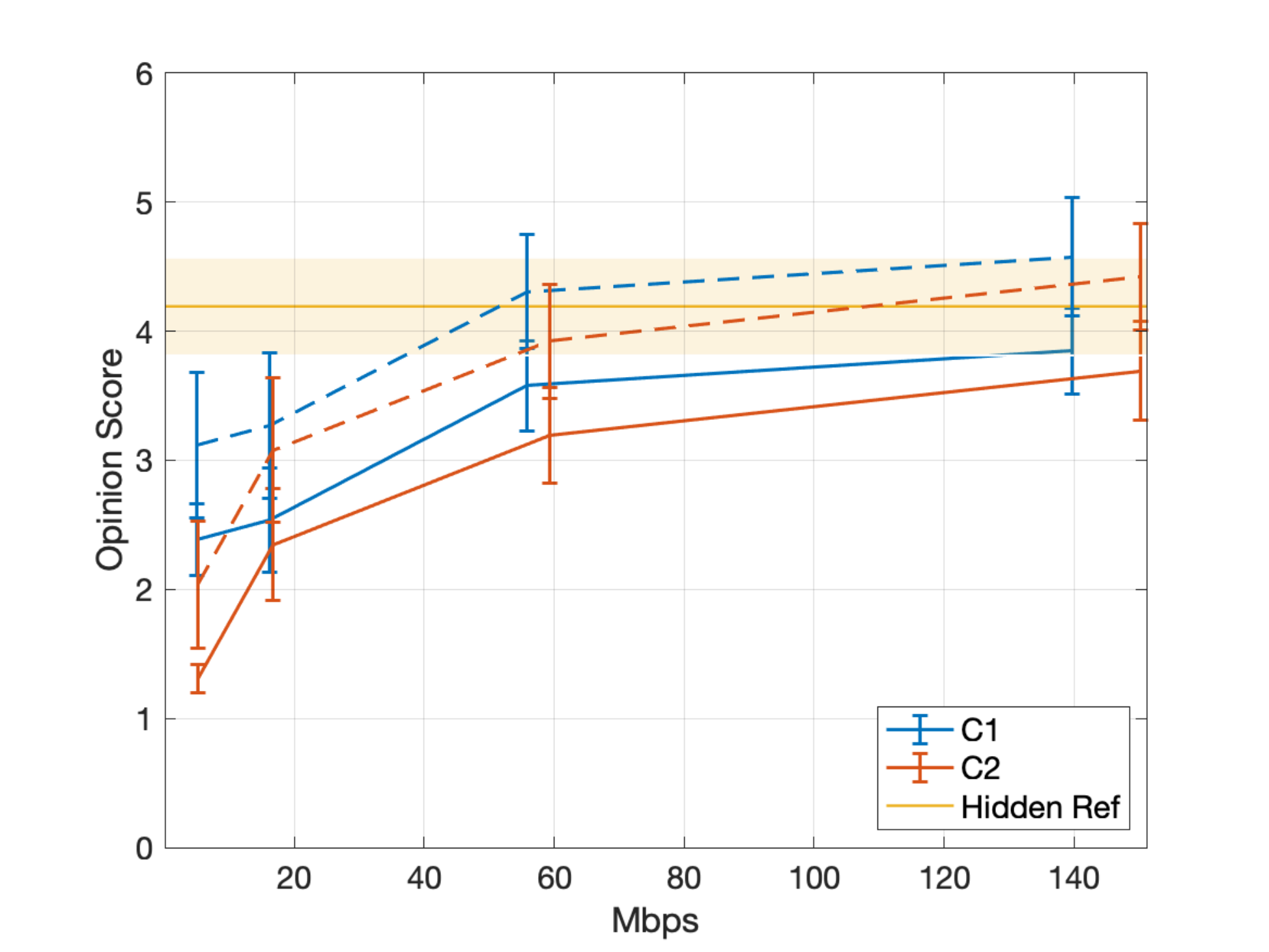}}\hf
\subfloat[\emph{Despoina}]{\includegraphics[width=\wtc]{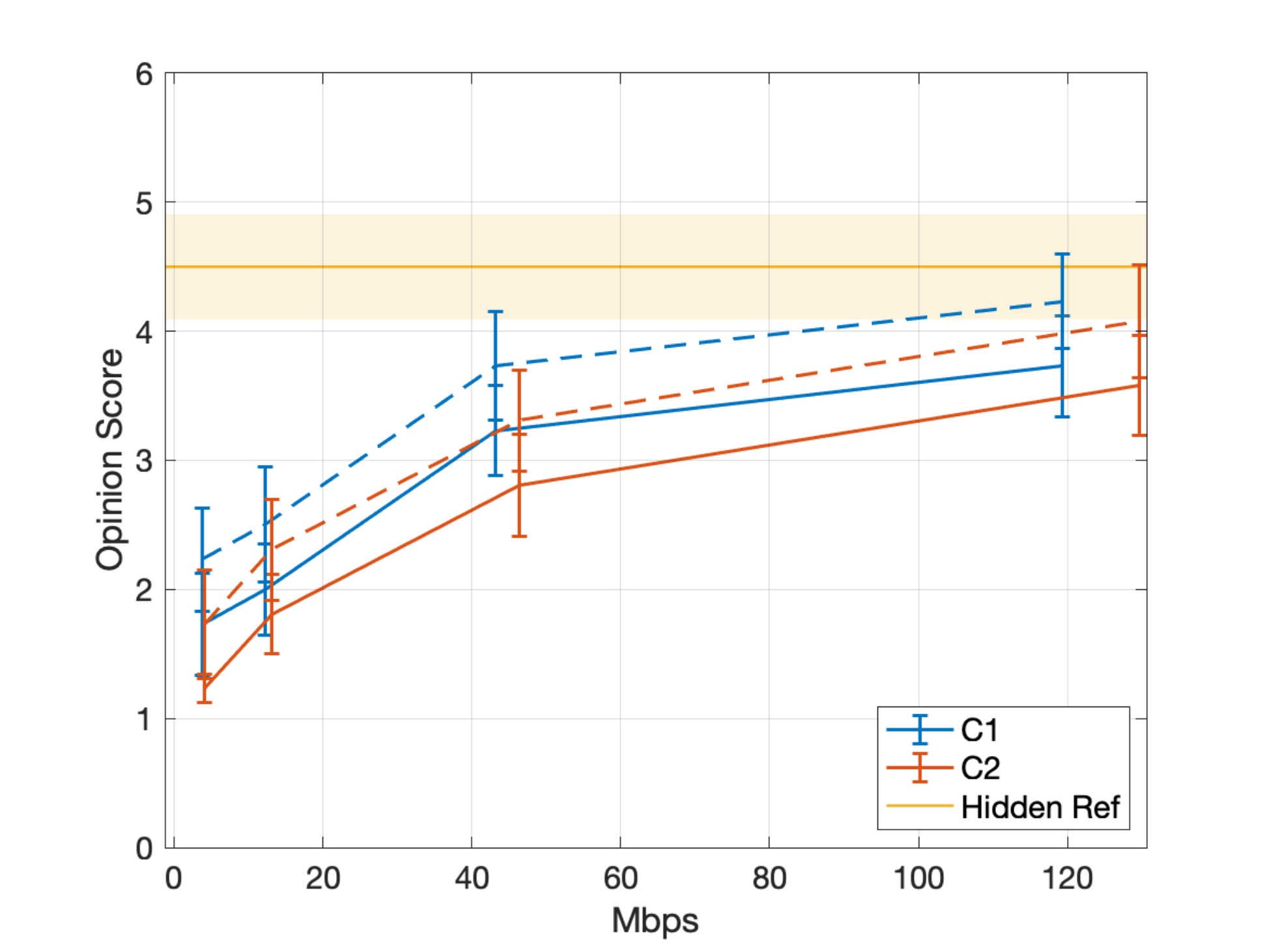}}\hf
\subfloat[\emph{Queen}]{\includegraphics[width=\wtc]{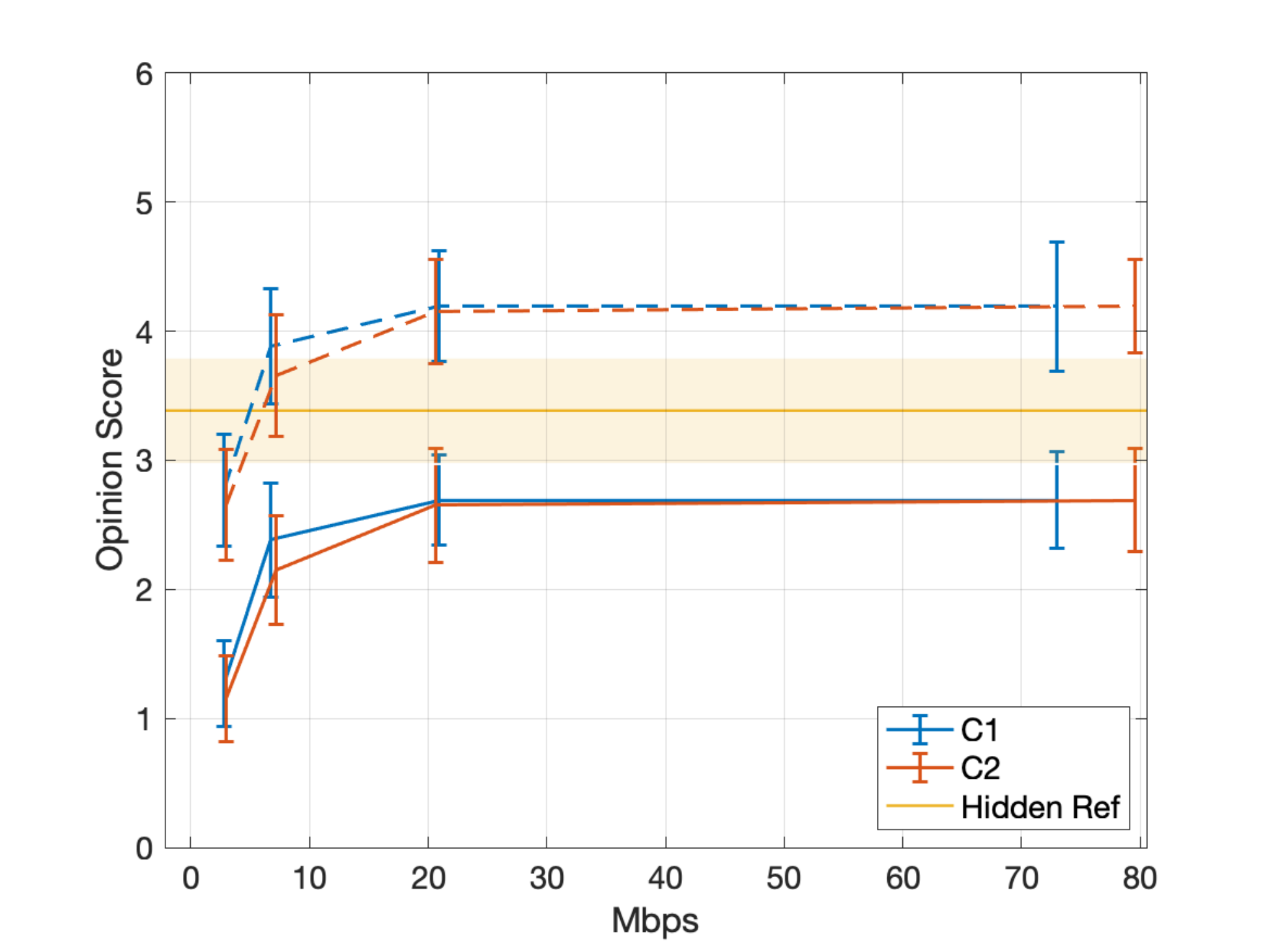}}\hf

\subfloat[\emph{Manfred}]{\includegraphics[width=\wtc]{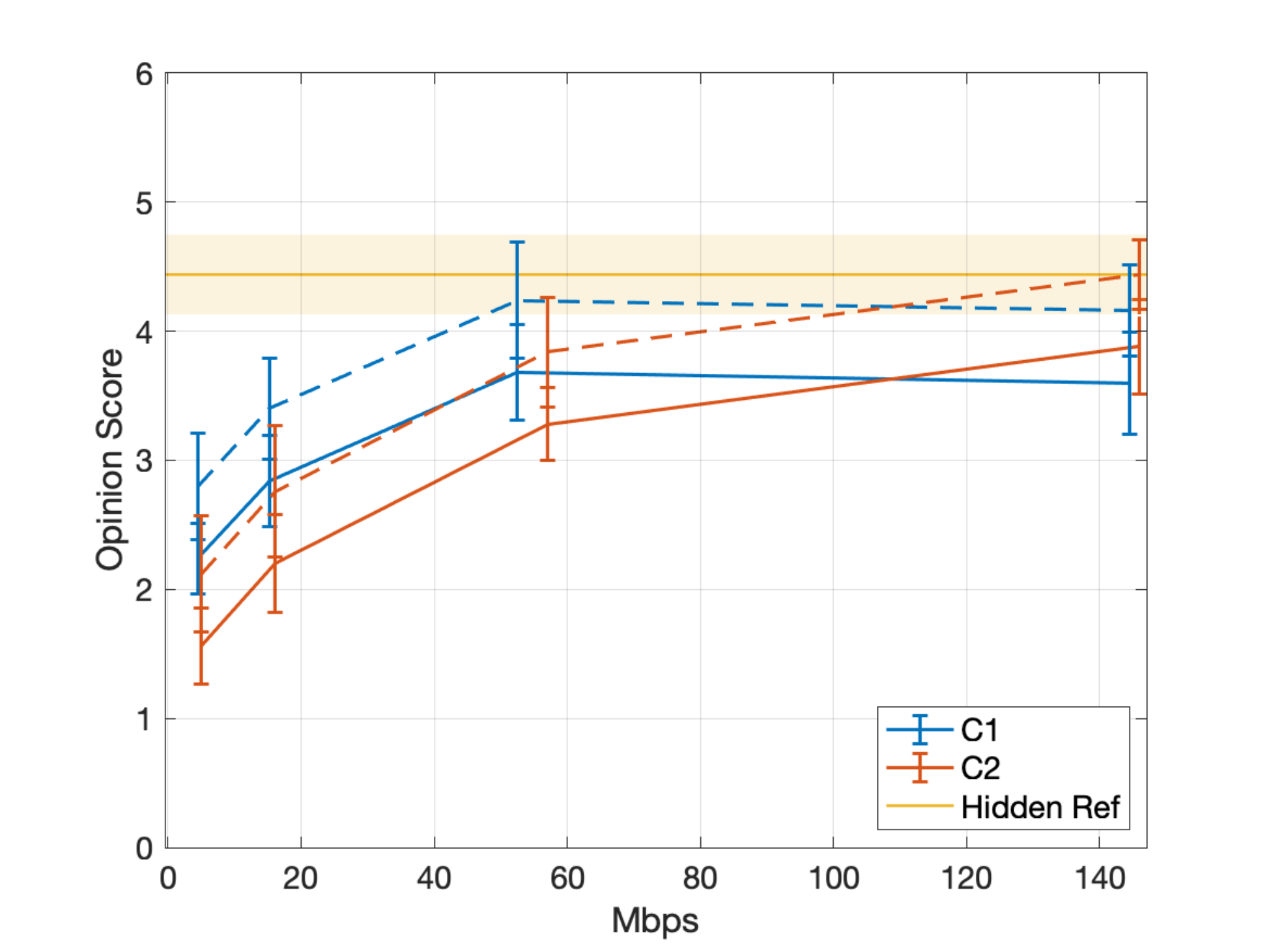}}\hf
\subfloat[\emph{Sarge}]{\includegraphics[width=\wtc]{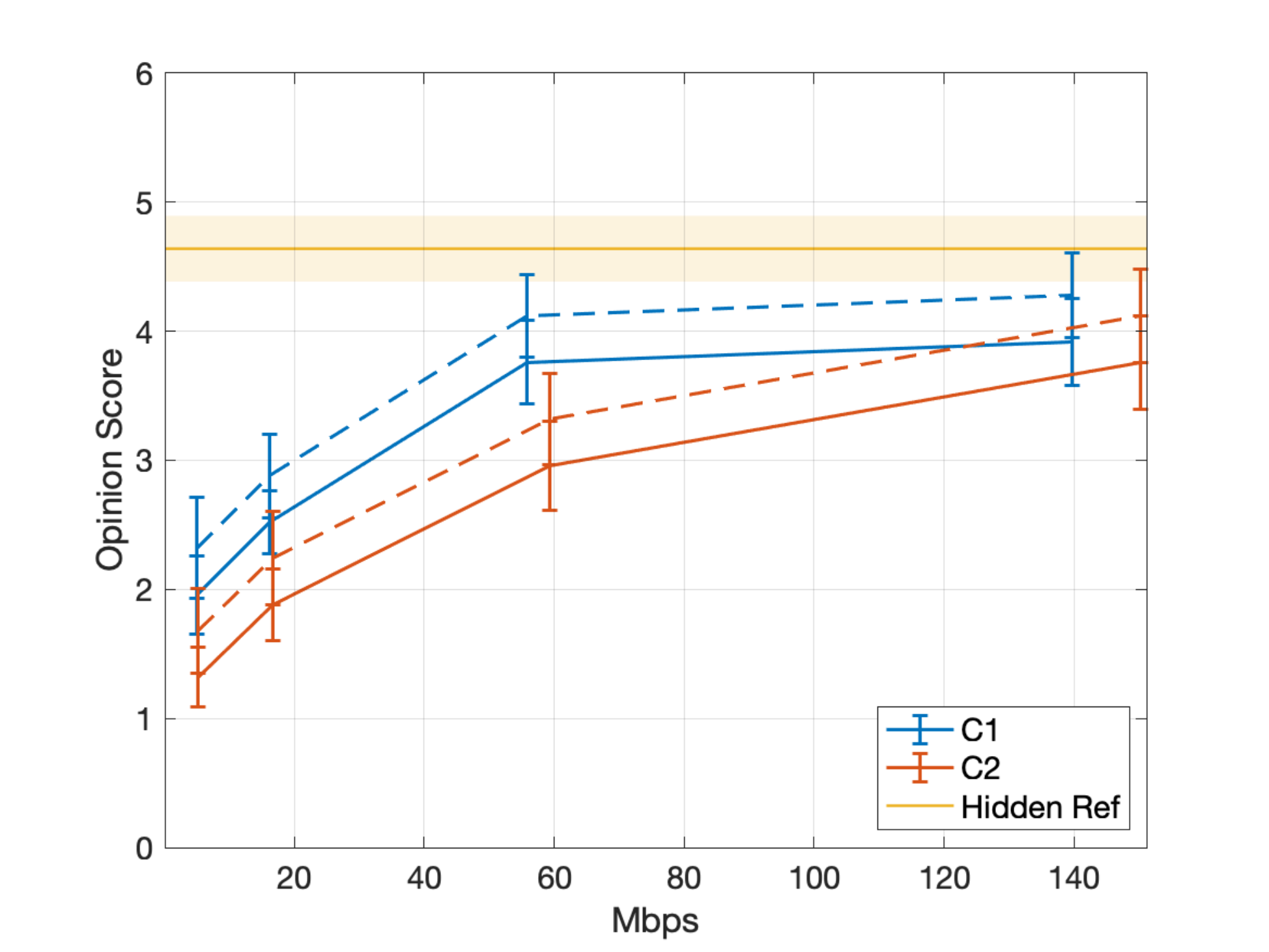}}\hf
\subfloat[\emph{Despoina}]{\includegraphics[width=\wtc]{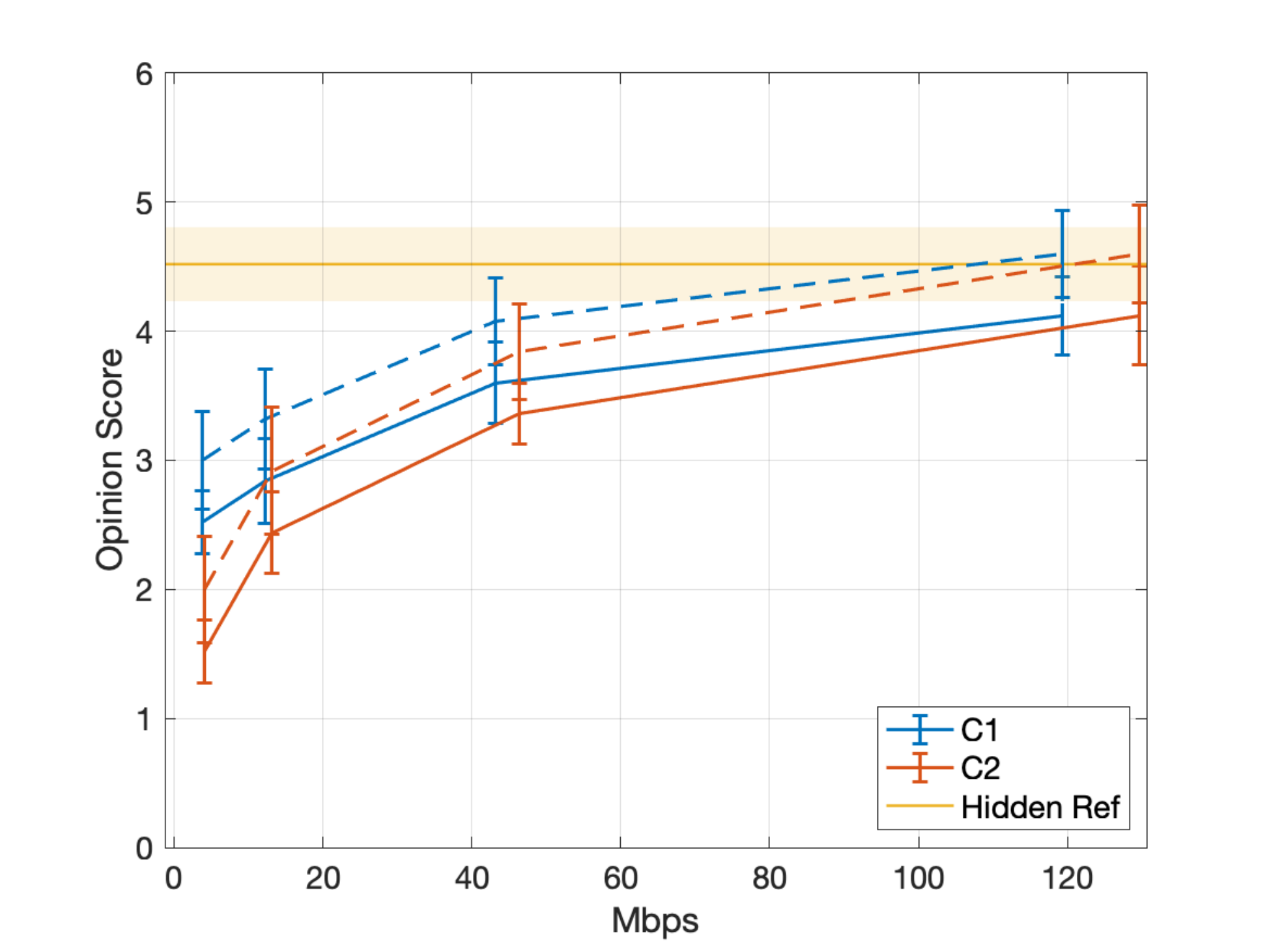}}\hf
\subfloat[\emph{Queen}]{\includegraphics[width=\wtc]{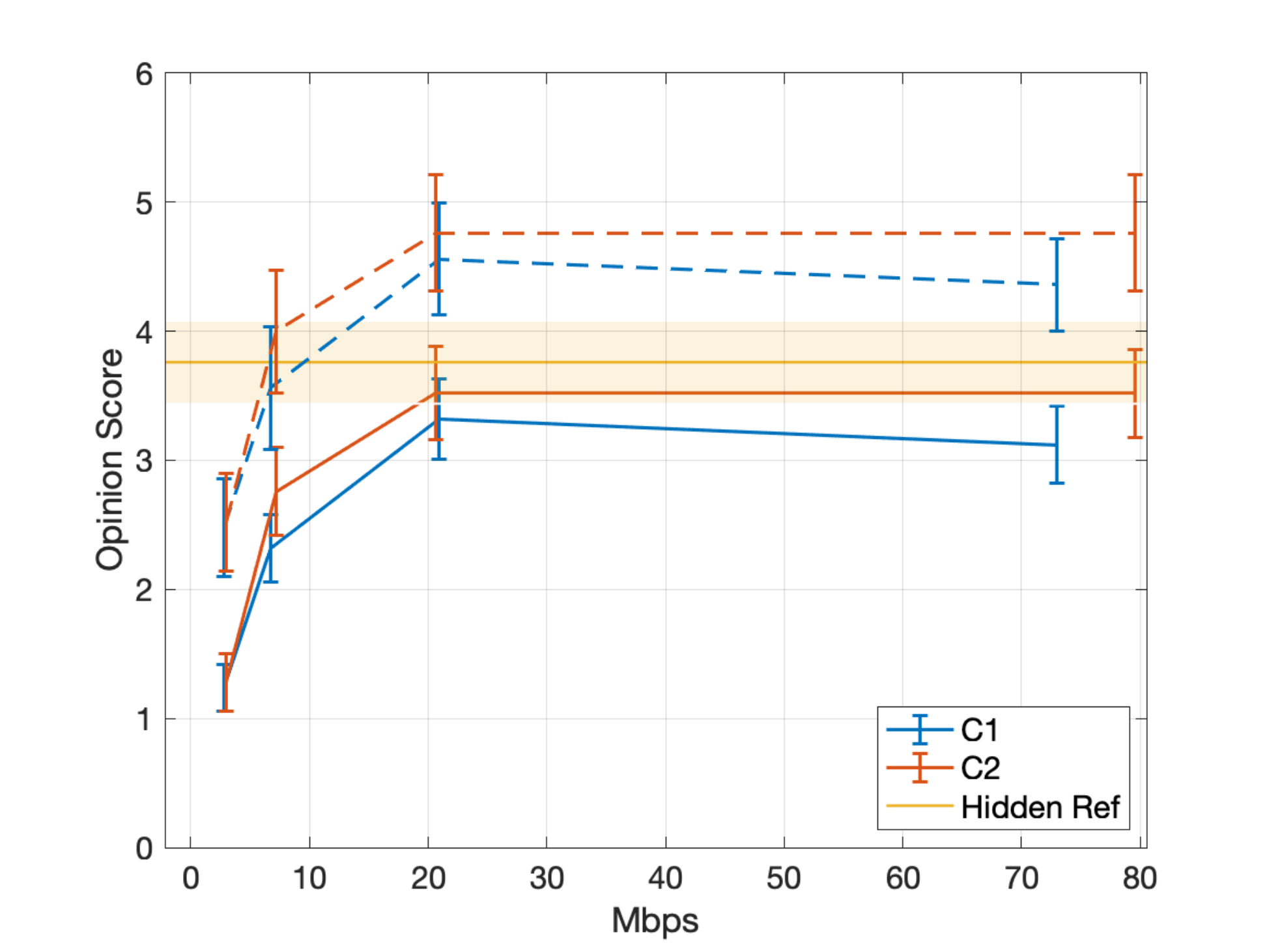}}\hf
\hf
\end{center}
\caption{{DMOS against achieved bit-rate. HR scores are shown using a dashed plot. Each column represents a content in test T1, whereas the rows depict results obtained using the viewing conditions 6DoF, 3DoF and 2DTV, respectively.}}
\label{fig:T1mos}
\vspace{-0.5cm}
\end{figure*}

\begin{figure*}[!t]
\begin{center}
\hf
\subfloat[\emph{Long dress}]{\includegraphics[width=\wtc]{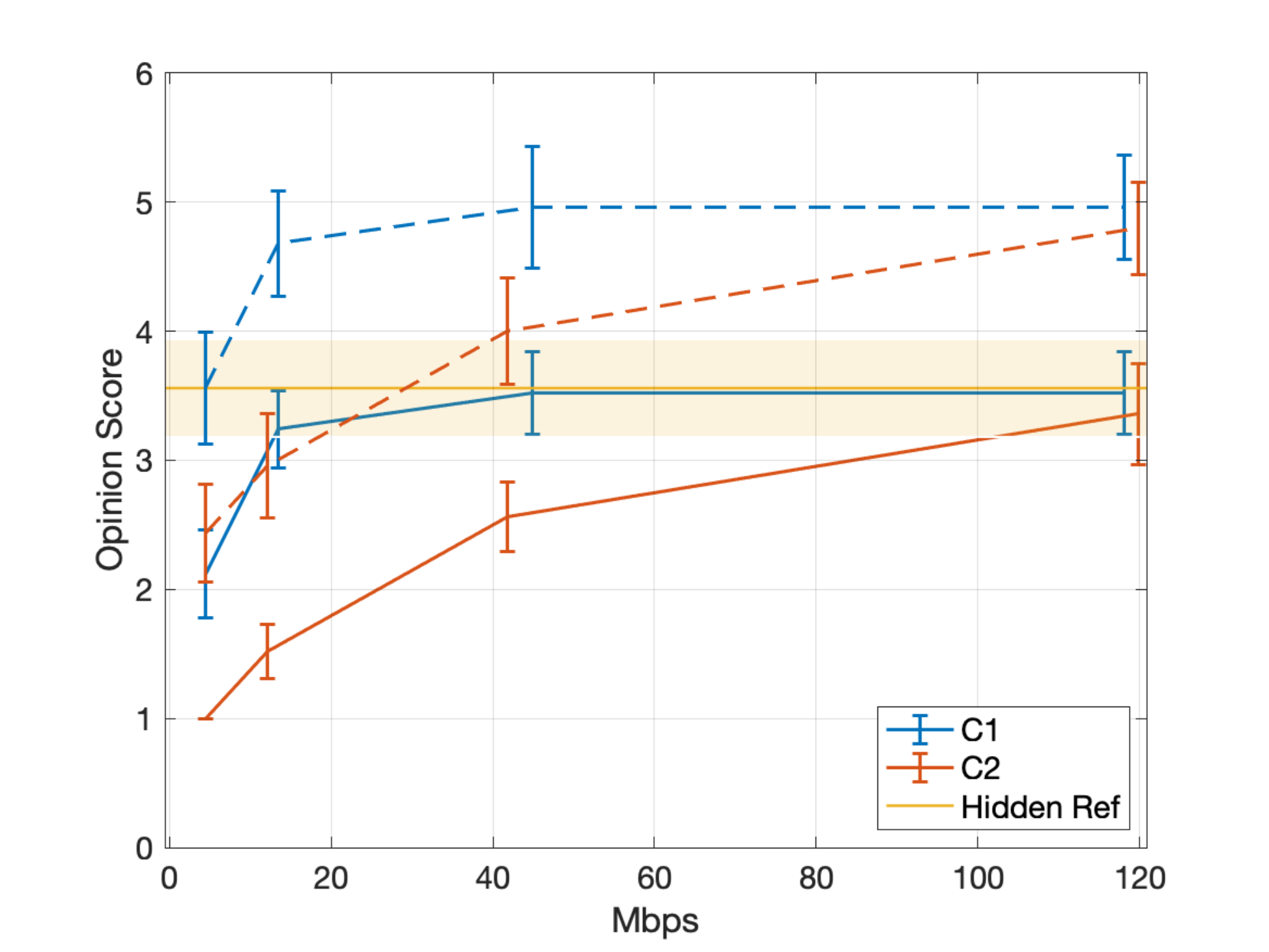}}\hf
\subfloat[\emph{Loot}]{\includegraphics[width=\wtc]{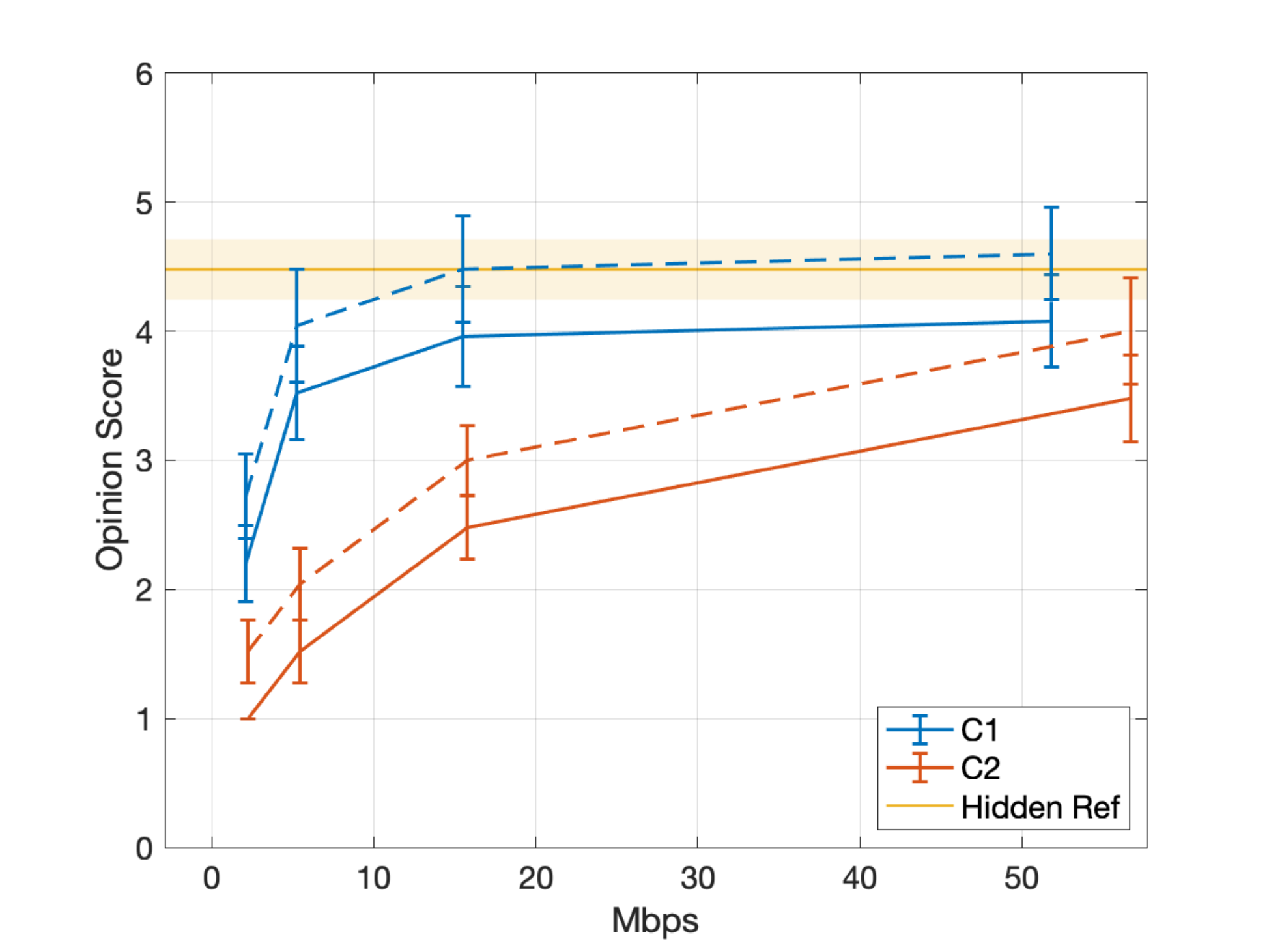}}\hf
\subfloat[\emph{Red and black}]{\includegraphics[width=\wtc]{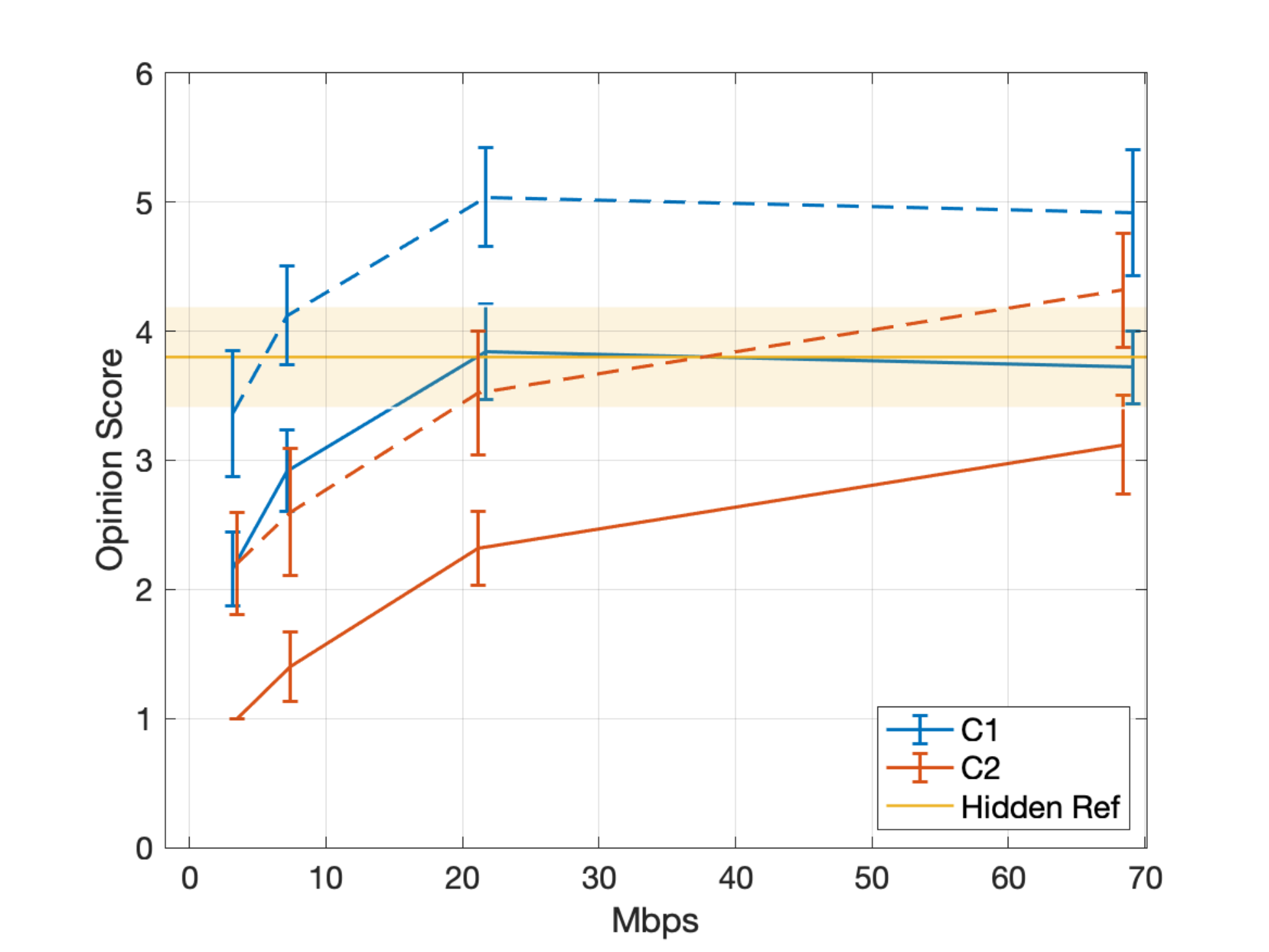}}\hf
\subfloat[\emph{Soldier}]{\includegraphics[width=\wtc]{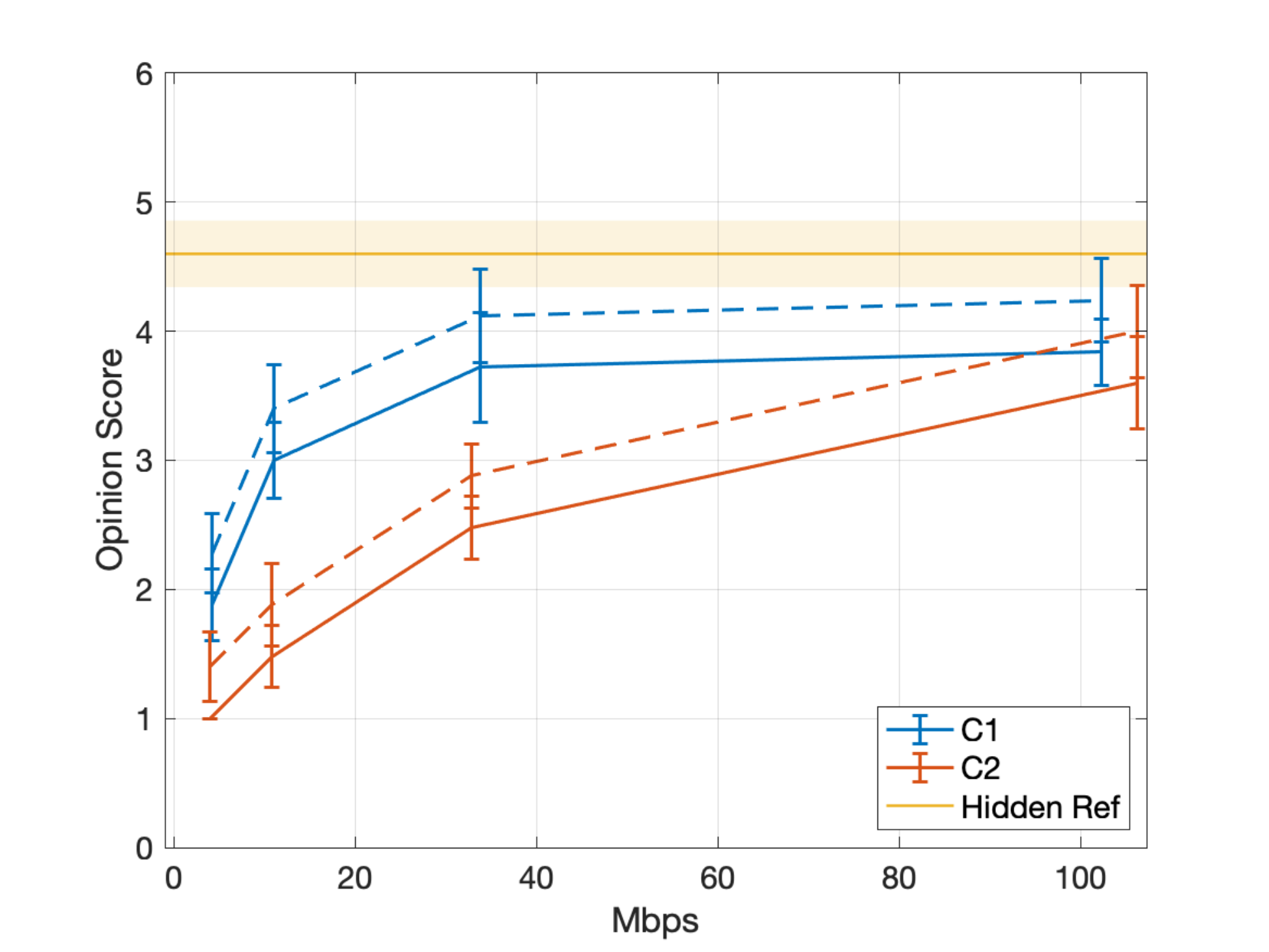}}\hf

\subfloat[\emph{Long dress}]{\includegraphics[width=\wtc]{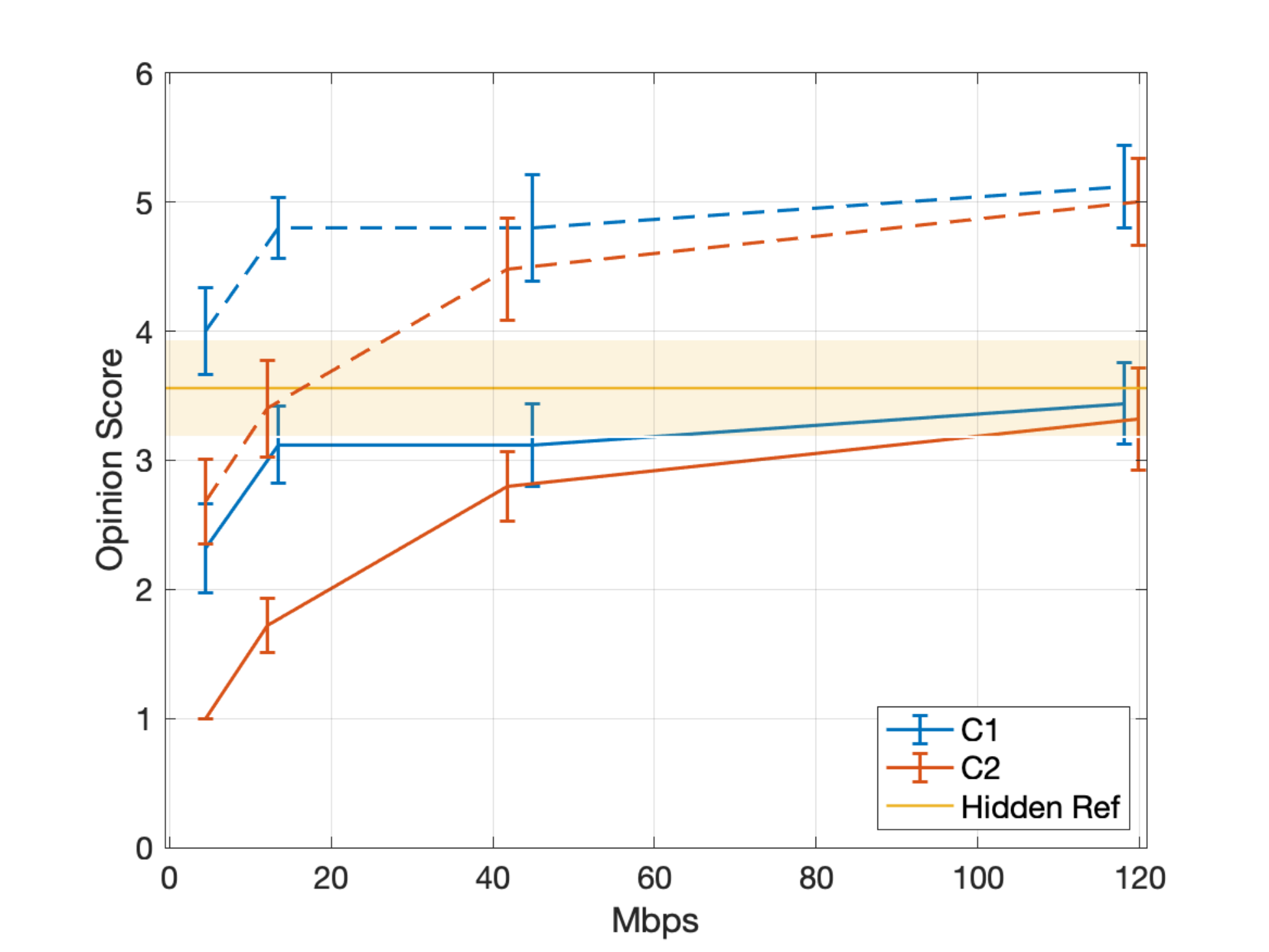}}\hf
\subfloat[\emph{Loot}]{\includegraphics[width=\wtc]{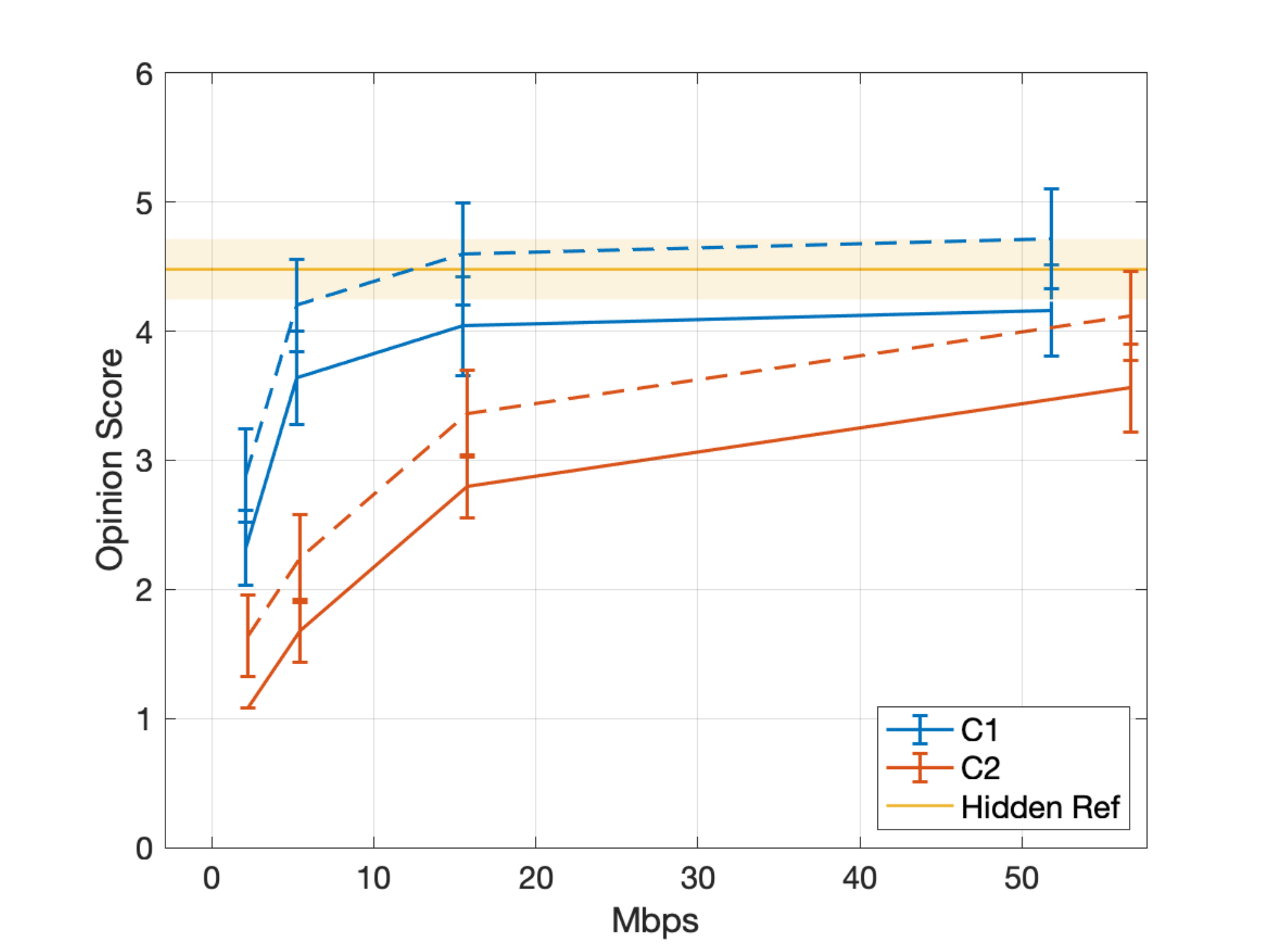}}\hf
\subfloat[\emph{Red and black}]{\includegraphics[width=\wtc]{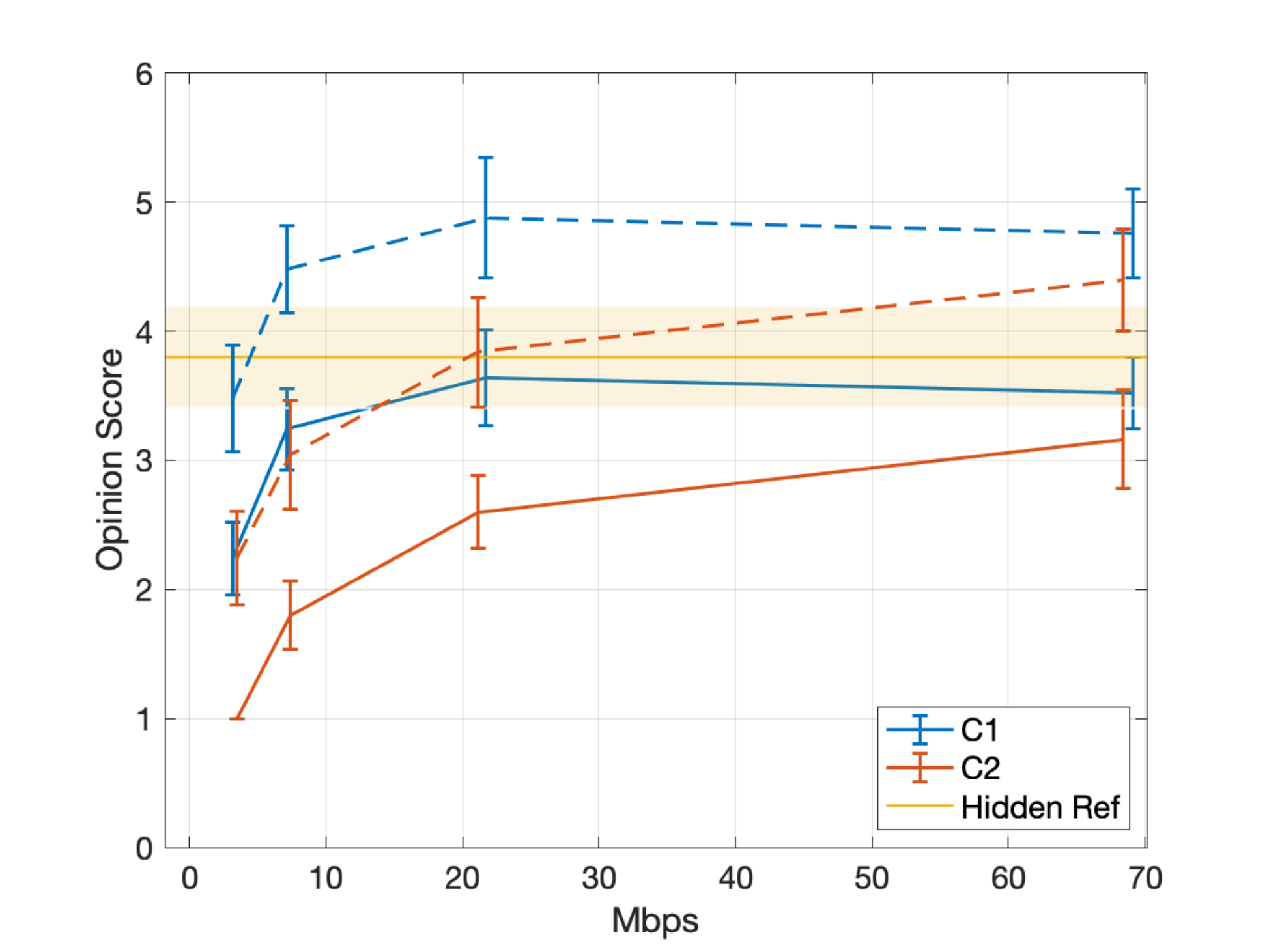}}\hf
\subfloat[\emph{Soldier}]{\includegraphics[width=\wtc]{_figures/soldier_6DoF-eps-converted-to.pdf}}\hf

\subfloat[\emph{Long dress}]{\includegraphics[width=\wtc]{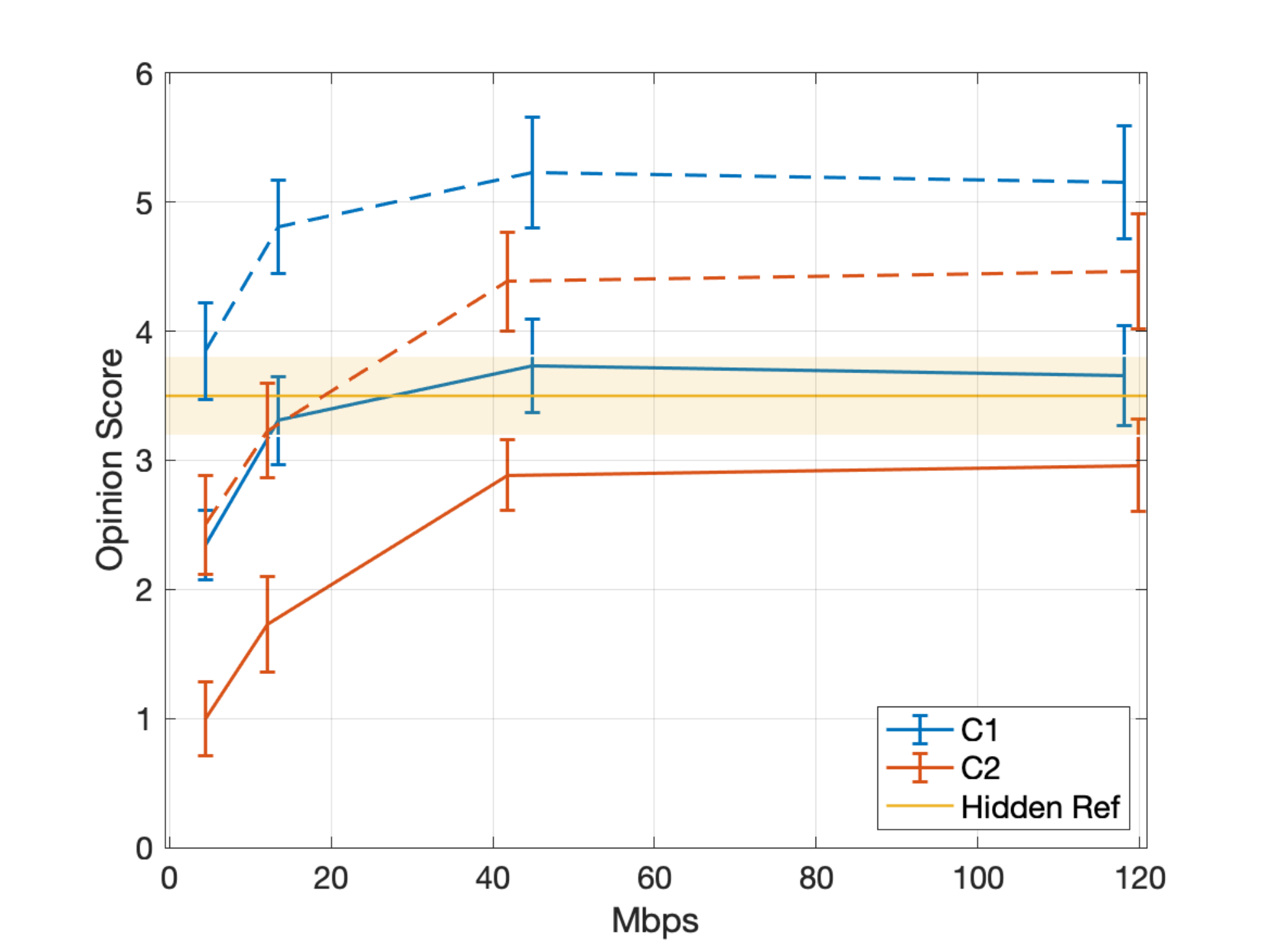}}\hf
\subfloat[\emph{Loot}]{\includegraphics[width=\wtc]{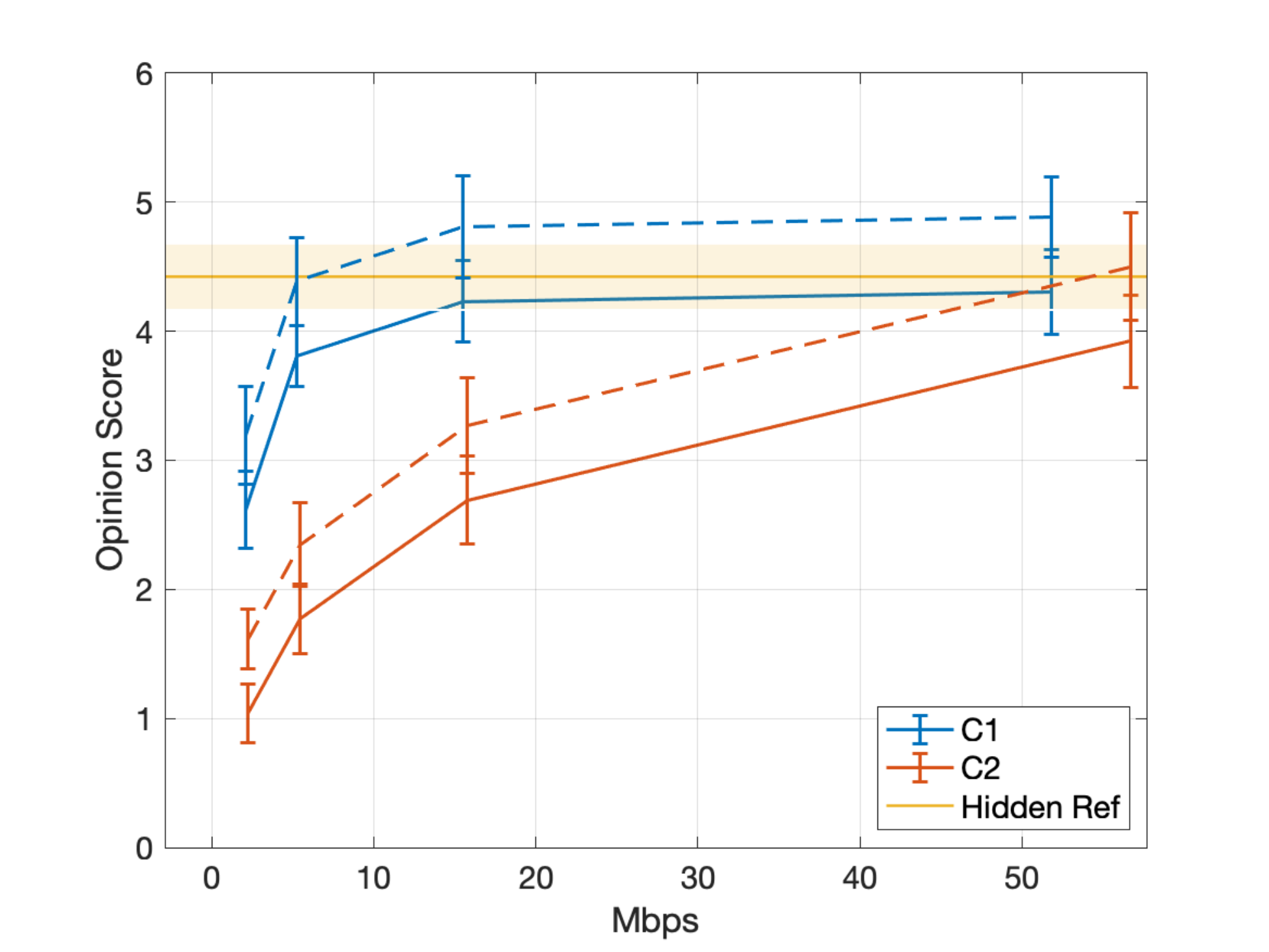}}\hf
\subfloat[\emph{Red and black}]{\includegraphics[width=\wtc]{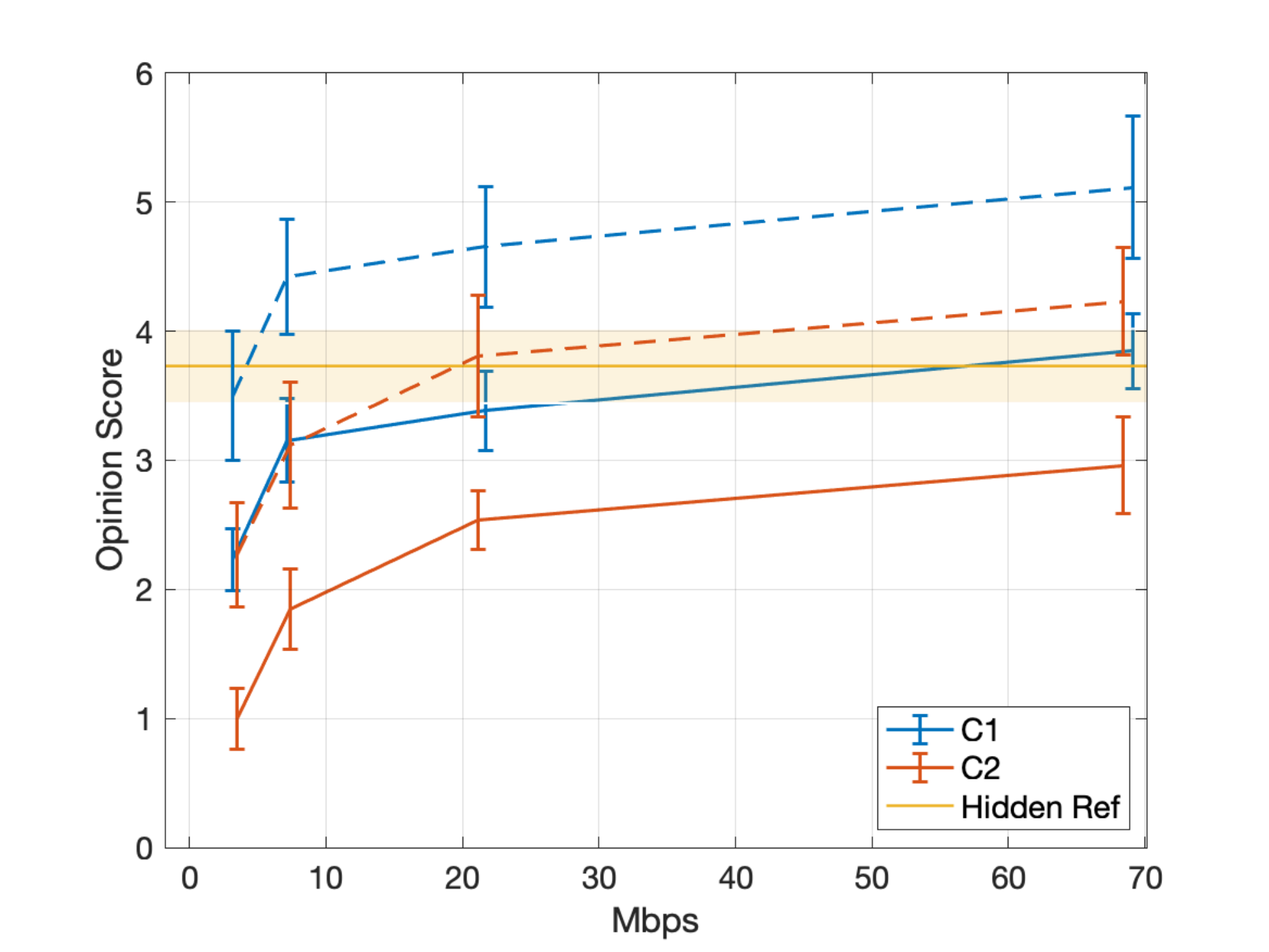}}\hf
\subfloat[\emph{Soldier}]{\includegraphics[width=\wtc]{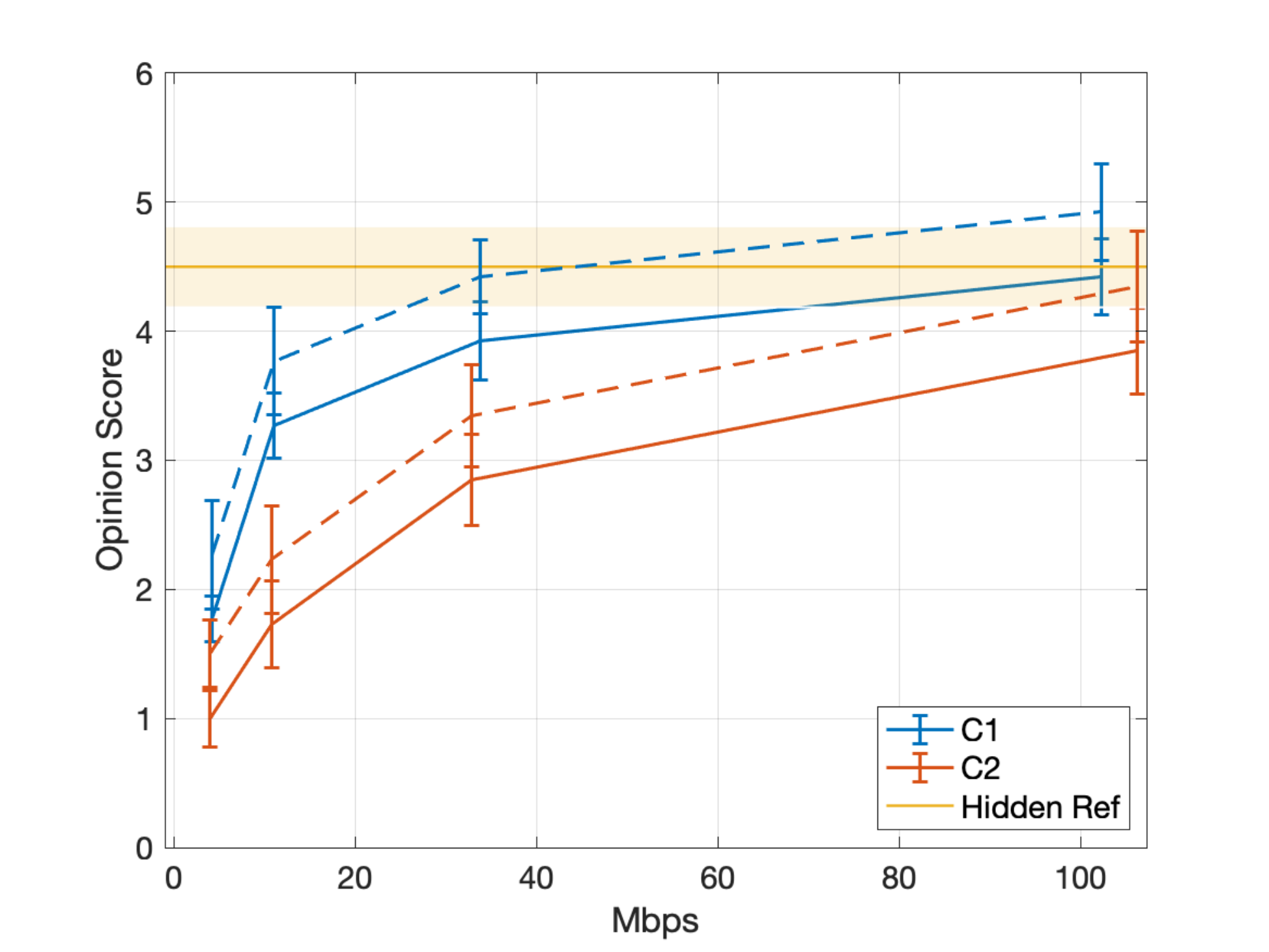}}\hf

\hf
\end{center}
\caption{{DMOS against achieved bit-rate. HR scores are shown using a dashed plot. Each column represents a content in test T2, whereas the rows depict results obtained using the viewing conditions 6DoF, 3DoF and 2DTV, respectively.}}
\label{fig:T2mos} 
\vspace{-0.5cm}
\end{figure*}
\section{Results}
\label{sec:results}

\color{black}
\subsection{Subjective quality assessment}
\label{subsec:qoe}
Figures~\ref{fig:T1mos} and ~\ref{fig:T2mos} shows the results of the subjective quality assessment of the contents comprising test T1 and test T2, for the 6DoF, 3DoF and 2DTV viewing conditions. In particular, the DMOS scores associated with the compressed contents are shown with solid lines, along with relative CIs, whereas the dashed lines represent the respective HR scores with the corresponding confidence intervals. 

To assess whether significant differences could be found between the three viewing conditions under test, we ran a non-parametric Kruskal-Wallis test, separately for test T1 and T2. The test was chosen as the gathered data was not found to be normally distributed, according to the Shapiro-Wilk normality test (T1: $W = 0.906$, $p < .001$; T2: $W = 0.909$, $p < .001$).
We found a significant effect of the viewing condition on the scores for test T1 ($\chi^2 = 37.56$, $p < .001$). Post-hoc analysis using Mann-Whitney U test with Bonferroni correction ($\alpha = .05/3$) revealed significant differences between 6DoF and 2DTV viewing conditions ($p < .001$, $r = 0.09$), and between 3DoF and 2DTV viewing conditions ($p < .001$, $r = 0.14$), but not between 6DoF and 3DoF ($p = 0.101$, $r = 0.04$). Values seem to indicate an effect of VR viewing condition with respect to traditional TV viewing on the final scores; however, the effect sizes suggest that, if the effect indeed exists, it is small. For test T2, no significant effect was found for viewing condition on the scores ($\chi^2 = 5.19$, $p = 0.075$).

It can be observed that codec \textit{C1} has generally a more favorable performance with respect to \textit{C2}. This is especially evident for the contents acquired through photogrammetry (see Fig.~\ref{fig:T2mos}), for which the gap among the two codecs is more pronounced. Mann-Whitney U test confirmed statistical significance for the two codecs (T1: $Z = 6.60$, $p < .001$, T2: $Z = 22.06$, $p < .001$), albeit with notably different effect sizes between test T1 and T2 ($r = 0.13$ and $r = 0.45$, respectively).

A Kruskal-Wallis  test performed on the scores revealed a significant effect of the content on the final scores, for both sets of contents
(T1: $\chi^2 = 64.91$, $p < .001$, T2: $\chi^2 = 35.23$, $p < .001$). Table~\ref{table:contents} shows the results of the post-hoc test conducted using Mann-Whitney U test with Bonferroni correction ($\alpha = .05/6$). Contents \textit{Manfred, Sarge} and \textit{Despoina} all show statistical significance with respect to content Queen ($p < .001$, $r \geq 0.15$ for all pairs). 
For contents acquired through photogrammetry, statistical significance was found between contents \textit{Longdress} and \textit{Loot}, and \textit{Loot} and \textit{Red and black} ($p < .001$, $r  = 0.14$ in both cases), as well as between contents \textit{Long dress} and \textit{Soldier} ($p = 0.006$, $r = 0.07$).
Results suggests that contents Long
dress and Red and black appear to be given different scores with respect to contents Loot and Soldier. However, the effect sizes suggest that the effect, if existent, is quite small.

We also ran a Kruskal-Wallis test on the scores to assess whether
the selected bit-rates were showing statistical significance. Results confirmed that the bit-rates have a significant effect for both tests
(T1: $\chi^2 = 1164.14$, $p < .001$, T2: $\chi^2 = 1008.42$, $p < .001$). Post-hoc analysis using Mann-Whitney U test with Bonferroni correction
($\alpha = .05/6$), shown in Table~\ref{table:rates} further confirmed that all pairwise
comparisons were statistically significant, for both test T1 and T2 ($p < .001, r > 0.20$ for all pairs).

\begin{table}[]

\centering
\caption{{Pairwise post-hoc test on the contents for test T1 and T2, using Wilcoxon signed-rank test with Bonferroni correction.}} 
\label{table:contents}
\vspace{-0.2cm}
\begin{tabular}{@{}cl|ccc@{}}
\toprule
           &                & $Z$    & $p$               & $r$    \\ \midrule
\multirow{6}{*}{\STAB{\rotatebox[origin=c]{90}{T1}}} &\textit{Manfred - Sarge}            & -1.10 & 0.270 & 0.03 \\
&\textit{Manfred - Despoina}         & -1.59 & 0.111           & 0.04 \\
&\textit{Manfred - Queen}            & 5.57 & \textless .001  & 0.15 \\
&\textit{Sarge - Despoina}           & -0.54 & 0.588           & 0.01 \\
&\textit{Sarge - Queen}              & 6.75 & \textless{}.001 & 0.18 \\
&\textit{Despoina - Queen}           & 7.04 & \textless{}.001 & 0.19 \\ \midrule
\multirow{6}{*}{\STAB{\rotatebox[origin=c]{90}{T2}}} &\textit{Long dress - Loot}          & -5.19 & \textless{}.001 & 0.14 \\
&\textit{Long dress - Red and black} & -0.05 & 0.960          & 0.00 \\
&\textit{Long dress - Soldier}       & -2.73 & 0.006 & 0.07 \\
&\textit{Loot - Red and black}       & 4.99 & \textless{}.001 & 0.14 \\
&\textit{Loot - Soldier}             & 2.08 & 0.037 & 0.06 \\
&\textit{Red and black - Soldier}    & -2.61 & 0.009           & 0.07 \\ \bottomrule
\end{tabular}
\vspace{-0.3cm}
\end{table}

\begin{table}[h]
\centering
\caption{{Pairwise post-hoc test on the bitrates for test T1 and T2, using Wilcoxon signed-rank test with Bonferroni correction.}}
\label{table:rates}
\vspace{-0.2cm}
\begin{tabular}{@{}cl|ccc@{}}
\toprule
&        & $Z$      & $p$               & $r$    \\ \midrule
\multirow{6}{*}{\STAB{\rotatebox[origin=c]{90}{T1}}} &R1 - R2 & -14.30 & \textless{}.001 & 0.41 \\
&R1 - R3 & -25.54 & \textless{}.001 & 0.73 \\
&R1 - R4 & -27.36 & \textless{}.001  & 0.78 \\
&R2 - R3 & -17.03 & \textless{}.001 & 0.49 \\
&R2 - R4 & -21.05 & \textless{}.001 & 0.60 \\
&R3 - R4 & -7.30  & \textless{}.001 & 0.21 \\ \midrule
\multirow{6}{*}{\STAB{\rotatebox[origin=c]{90}{T2}}} &R1 - R2 & -14.61 & \textless{}.001 & 0.42 \\
&R1 - R3 & -23.84 & \textless{}.001 & 0.68 \\
&R1 - R4 & -27.39 & \textless{}.001 & 0.79 \\
&R2 - R3 & -11.20 & \textless{}.001 & 0.32 \\
&R2 - R4 & -17.95 & \textless{}.001 & 0.51 \\
&R3 - R4 & -8.53  & \textless{}.001 & 0.24 \\ \bottomrule
\end{tabular}
\vspace{-0.5cm}
\end{table}

\subsection{Data analysis}
\subsubsection{T1}
 In order to further analyze the effect of DoF conditions, contents, codecs and bit-rates, and relative interactions, on the gathered scores, we fitted a full linear mixed-effects model on the data, accounting for randomness introduced by the participants. 
Due to the non-normality of our data, the aligned rank transform was applied prior to the fitting~\cite{wobbrock2011aligned}. Since the transform is designed for a fully randomized test, it is not suitable for the scores collected during the test, as the HR addition makes the design matrix rank deficient. However, the transform can be applied to the differential scores used to obtain DMOS, as it follows a fully randomized design. Thus, it was decided to perform the analysis on the differential scores. Post-hoc contrast tests were performed using the ART-C tool~\cite{elkin2021aligned}.


\begin{table}[]
\caption{Analysis of deviance on the full mixed-effects model, for test T1.}
\label{table:ANOVA1}
\vspace{-0.2cm}
\centering
\begin{tabular}{lrrr}
\toprule
&$F$&$df$&$p$ \\ \midrule
DoF&                            $0.07$&     $2$&        $ 0.936$ \\
Content&                        $53.98$&    $3$&        $ <.001$ \\
Codec&                          $65.19$&    $1$&        $ <.001$ \\
Bitrate&                        $595.38$&   $3$&        $ <.001$ \\
DoF: Content&                   $11.39$&    $6$&        $ <.001$ \\
DoF: Codec&                     $0.58$&     $2$&        $ 0.660$ \\
Content: Codec&                 $9.97$&     $3$&        $ <.001$ \\
DoF: Bitrate&                   $1.70$&     $6$&        $ 0.116$ \\
Content: Bitrate&               $6.04$&     $9$&        $ <.001$ \\
Codec: Bitrate&                 $8.96$&     $3$&        $ <.001$ \\
DoF: Content: Codec&            $0.71$&     $6$&        $ 0.643$ \\
DoF: Content: Bitrate&          $0.98$&     $18$&       $ 0.480$ \\
DoF: Codec: Bitrate&            $0.73$&     $6$&        $ 0.627$ \\
Content: Codec: Bitrate&        $1.23$&     $9$&        $ 0.273$ \\
DoF: Content: Codec: Bitrate&   $0.48$&     $18$&       $ 0.969$ \\ \bottomrule
\hline
\end{tabular}
\vspace{-0.3cm}
\end{table}

For test T1, analysis of deviance on the full mixed-effects model showed significance for main effects Content ($F = 53.98$, $df = 3$, $p < .001$), Codec ($F = 65.19$, $df = 1$, $p < .001$) and bit-rate ($F = 595.38$, $df = 3$, $p < .001$), but not for DoF ($F = 0.07$, $df = 2$, $p = 0.936$). Moreover, significant interaction effects were found for \textit{DoF - Content} ($F = 11.39$, $df = 6$, $p < .001$), \textit{Content - Codec} ($F = 9.97$, $df = 3$, $p < .001$), \textit{Content - bit-rate} ($F = 6.04$, $df = 9$, $p < .001$) and \textit{Codec - bit-rate} ($F = 8.96$, $df = 3$, $p < .001$). No interaction effect beyond the first level was found to be significant. The full results of the analysis of deviance can be found in Table~\ref{table:ANOVA1}. 

Post-hoc interaction analysis using ART-C revealed significant differences at $5\%$ level in 3DoF for content pairs \textit{Manfred - Queen} ($p < .001$), \textit{Sarge - Queen} ($p < .001$) and \textit{Despoina - Queen} ($p < .001$); in 6DoF, for content pairs \textit{Manfred - Sarge} ($p < .001$) and \textit{Despoina - Queen} ($p < .001$); in the 2DTV condition, for content pairs \textit{Manfred - Sarge} ($p = 0.019$) \textit{Manfred - Queen} ($p < .001$), \textit{Sarge - Despoina} ($p < .001$), \textit{Sarge - Queen} ($p < .001$), and \textit{Despoina - Queen} ($p = 0.024$). Additionally, several content pairs exhibited significant contrasts when different DoF were employed; for a full report of the contrasts, we invite readers to consult the appendix. Most notably, no significant effect was found when the same content was displayed in different DoF mediums; that is, for every content under exam, the pairs \textit{3DoF - 6DoF, 3DoF - 2DTV}, and \textit{6DoF - 2DTV} were consistently above the $5\%$ threshold of significance.

Regarding the interaction between factors Content and Codec, post-hoc analysis using ART-C showed significant differences at $5\%$ level between the two codecs for all contents ($p < .001$) except \textit{Queen}, for which the two codecs were deemed equivalent ($p = 0.995$). Furthermore, statistically significant differences were found, for \textit{C1}, in content pairs \textit{Manfred - Despoina} ($p = 0.006$), and \textit{Despoina - Queen} ($p < .001$), and for \textit{C2}, in content pairs \textit{Manfred - Despoina} ($p = 0.010$), \textit{Manfred - Queen} ($p < .001$), \textit{Sarge - Queen} ($p < .001$), and \textit{Despoina - Queen} ($p < .001$). Several content pairs were found to have significant contrasts when different codecs were employed; for a complete overview, we refer readers to the appendix.

Significant differences at $5\%$ level were also found when considering post-hoc interactions between factors Content and Bitrate. In particular, the pair \textit{Manfred - Queen} was found to have significant differences for rate \textit{R2} ($p < .001$) and \textit{R3} ($p = 0.018$); pair \textit{Sarge - Queen} for rate \textit{R1} ($p = 0.033$), \textit{R2} ($p < .001$), and \textit{R3} ($p = 0.005$); pair \textit{Despoina - Queen}, for rate \textit{R1} ($p < .001$), \textit{R2} ($p < .001$), and \textit{R3} ($p < .001$). No significant difference was found among contents at rate \textit{R4}, indicating that, at the highest quality level, the contents were rated similarly. As expected, most of the comparisons between different bitrates, be it with the same or among different contents, yield statistically significant differences; the complete results are available in the appendix.     
Finally, results of the post-hoc analysis of interactions between factors Codec and Bitrate showed statistically significant differences at $5\%$ level among the codecs for rate \textit{R1} ($p < .001$) and \textit{R3} ($p = 0.007$), but not for rate \textit{R2} ($p = 0.052$ and \textit{R4}) ($p = 0.986$). In the first case, the p-value is quite close to significance, whereas in the second, the test confirms our previous observations: at the highest quality level, the difference among codecs seems to be imperceptible. The rest of the combinations between codecs and bitrates lead to significant differences, with the exception of \textit{C1-R3} versus \textit{C2-R4} ($p = 0.054$), indicating that the null hypothesis cannot be rejected for \textit{C2} at rate \textit{R4} when compared to \textit{C1} at rate \textit{R3}; in other words, to achieve similar ratings to codec \textit{C1} (MPEG V-PCC), codec \textit{C2} (the MPEG anchor) requires higher bandwidth, which is in line with what observed in the previous section. More exhaustive results can be found in the appendix.

\subsubsection{T2}

\begin{table}[]
\caption{Analysis of deviance on the full mixed-effects model, for test T2.}
\label{table:ANOVA2}
\vspace{-0.2cm}
\centering
\begin{tabular}{lrrr}
\toprule
&$F$&$df$&$p$ \\ \midrule
DoF&                            $2.59$&     $2$&        $ 0.082$ \\
Content&                        $165.56$&   $3$&        $ <.001$ \\
Codec&                          $1059.81$&  $1$&        $ <.001$ \\
Bitrate&                        $702.54$&   $3$&        $ <.001$ \\
DoF: Content&                   $1.99$&     $6$&        $ 0.064$ \\
DoF: Codec&                     $1.08$&     $2$&        $ 0.340$ \\
Content: Codec&                 $5.81$&     $3$&        $ <.001$ \\
DoF: Bitrate&                   $0.39$&     $6$&        $ 0.881$ \\
Content: Bitrate&               $6.30$&     $9$&        $ <.001$ \\
Codec: Bitrate&                 $44.89$&    $3$&        $ <.001$ \\
DoF: Content: Codec&            $0.80$&     $6$&        $ 0.569$ \\
DoF: Content: Bitrate&          $1.10$&     $18$&       $ 0.341$ \\
DoF: Codec: Bitrate&            $1.12$&     $6$&        $ 0.346$ \\
Content: Codec: Bitrate&        $2.08$&     $9$&        $ 0.028$ \\
DoF: Content: Codec: Bitrate&   $0.69$&     $18$&       $ 0.820$ \\ \bottomrule
\hline
\end{tabular}
\vspace{-0.3cm}
\end{table}
Results of analysis of deviance on the full mixed-effects model for test T2 showed significance for main effects Content ($F = 165.56$, $df = 3$, $p < .001$), Codec ($F = 1059.81$, $df = 1$, $p < .001$) and Bitrate ($F = 702.54$, $df = 3$, $p < .001$) but not for DoF ($F = 2.59$, $df = 2$, $p = 0.0825$), similarly to what was seen for test T1. 
Two-way interactions were found significant at 5\% level between Content and Codec ($F = 5.81$, $df = 3$, $p < .001$), Content and Bitrate ($F = 6.30$, $df = 9$, $p < .001$), and Codec and Bitrate ($F = 44.89$, $df = 3$, $p < .001$), as well as the three-way interaction between Content, Codec, and Bitrate ($F = 2.08$, $df = 9$, $p = 0.028$). 

Post-hoc interaction analysis using ART-C for the three-way interaction between factors Content, Codec, and Bitrate revealed significant differences at $5\%$ level between the two codecs under exam, when fixing content and bitrate level, for touples involving content \textit{Longdress} at bitrate \emph{R1} ($p < .001$), \textit{R2} ($p < .001$), and \textit{R3} ($p < .001$), but not for the highest bitrate \textit{R4}  ($p = 0.964$). Similarly, for content Soldier, touples had significant interactions at rate \textit{R1} ($p = 0.010$), \textit{R2} ($p < .001$), and  \textit{R3} ($p < .001$), but not \textit{R4} ($p = 0.085$). For the other two contents, all touples at same bitrate were significant (\textit{Loot, C1, R1 - Loot, C2, R1}: $p < .001$; \textit{Loot, C1, R2 - Loot, C2, R2}: $p < .001$;
\textit{Loot, C1, R3 - Loot, C2, R3}: $p < .001$; \textit{Loot, C1, R4 - Loot, C2, R4}: $p = 0.016$; \textit{Red and black, C1, R1 - Red and black, C2, R1}: $p < .001$; \textit{Red and black, C1, R2 - Red and black, C2, R2}: $p < .001$; \textit{Red and black, C1, R3 - Red and black, C2, R3}: $p < .001$; \textit{Red and black, C1, R4 - Red and black, C2, R4}: $p = 0.047$). This indicates that, with the exception of \textit{Longdress} and \textit{Soldier} at bitrate \textit{R4}, for all rate points and all contents the two codecs were significantly different at $5\%$ level.

Post-hoc interaction also revealed statistical difference among different contents: considering codec \textit{C1} at bitrate \textit{R1}, significant differences were found in content touples \textit{Longdress - Loot} ($p < .001$), \textit{Longdress - Soldier} ($p < .001$), \textit{Loot - Soldier} ($p = 0.004$), and \textit{Red and black - Soldier} ($p < .001$), but not for touples \textit{Longdress - Red and black} ($p = 0.626$) and \textit{Loot - Red and black} ($p = 0.091$); analogously, at bitrate \textit{R2}, differences were found for touples \textit{Longdress - Loot} ($p = 0.004$), \textit{Longdress - Soldier} ($p < .001$), \textit{Loot - Soldier} ($p < .001$), and \textit{Red and black - Soldier} ($p < .001$), but not for touples \textit{Longdress - Red and black} ($p = 0.198$) and \textit{Loot - Red and black} ($p = 1$).
At bitrate \textit{R3}, significant differences were found only between touples \textit{Longdress - Soldier} ($p < .001$); at bitrate \textit{R4}, all content touples failed to reject the null hypothesis. This indicates that differences in DMOS scores among different contents are more prominent at lowest bitrates for codec \textit{C1}, whereas for higher bitrates, contents received similar scores.
When considering codec \textit{C2}, at bitrate \textit{R1}, significant differences were observed for content touples \textit{Longdress - Loot} ($p < .001$), \textit{Longdress - Soldier} ($p < .001$), \textit{Loot - Red and black} ($p = 0.047$), and \textit{Red and black - Soldier} ($p = 0.004$), but not for touples \textit{Longdress - Red and black} ($p = 0.999$) and \textit{Loot - Soldier} ($p = 1$); similarly, at bitrate \textit{R2}, differences were deemed significant for content touples \textit{Longdress - Loot} ($p < .001$), \textit{Longdress - Soldier} ($p < .001$), \textit{Loot - Red and black}, ($p < .001$), and \textit{Red and black - Soldier} ($p < .001$), but not for \textit{Longdress - Red and black} ($p = 0.994$) and \textit{Loot - Soldier} ($p = 1$). This seems to indicate that, a lower bitrates, contents presenting the same gender were rated similarly. At bitrate \textit{R3}, statistical difference could be observed among touples \textit{Longdress - Loot} ($p < .001$), \textit{Longdress - Red and black} ($p = 0.018$), \textit{Longdress - Soldier} ($p < .001$), \textit{Loot - Red and black} ($p = 0.009$), and \textit{Red and black - Soldier} ($p < .001$), but not for touple \textit{Loot - Soldier} ($p = 1$), whereas for bitrate \textit{R4}, statistical differences at a significant level were only observed between \textit{Longdress - Soldier} ($p = 0.015$). Results indicate that for the highest bitrate, similar trends can be generally observed between different contents for both codecs under consideration, whereas as bitrate decreases, more differences can be spotted in DMOS trends. 

Finally, post-hoc interaction analysis revealed significant differences at $5\%$ level, for content \textit{Longdress} encoded with codec \textit{C1}, between bitrate \textit{R1} with respect with all the other bitrates (\textit{R1 - R2}: $p < .001$; \textit{R1 - R3}: $p < .001$; \textit{R1 - R4}: $p < .001$); however, no significant difference in rating was found when comparing \textit{R2}, \textit{R3}, and \textit{R4}. Remarkably, different trends can be observed for codec \textit{C2}, for which only bitrates \textit{R3 - R4} are to be considered statistically equivalent ($p = 0.055$), whereas all other touples for content \textit{Longdress} present significant differences (\textit{R1 - R2}: $p = 0.005$; \textit{R1 - R3}: $p < .001$; \textit{R1 - R4}: $p < .001$; \textit{R2 - R3}: $p < .001$; \textit{R2 - R4}: $p < .001$). For content \textit{Loot}, statistical differences are observed, for codec \textit{C1}, among all bitrates, bar \textit{R2 - R3} ($p = 0.142$) and \textit{R3 - R4} ($p = 1$); the rest falls below the significance level (\textit{R1 - R2}: $p < .001$; \textit{R1 - R3}: $p < .001$; \textit{R1 - R4}: $p < .001$; \textit{R2 - R4}: $p = 0.004$). For codec \textit{C2}, however, a different trend is observed: bitrates \textit{R1 - R2} are the only touple, for \textit{Loot}, for which no statistical difference is found ($p = 0.377$), whereas all other cases present statistically significant differences ($p < .001$ for all touples). Content \textit{Red and black} exhibits similar behaviour as \textit{Longdress}: for codec \textit{C1}, only differences among bitrate \textit{R1} with all the other bitrates are significant (\textit{R1 - R2}: $p < .001$; \textit{R1 - R3}: $p < .001$; \textit{R1 - R4}: $p < .001$), whereas for codec \textit{C2}, all bitrate touples are statistically different at $5\%$ significance level (\textit{R1 - R2}: $p = 0.018$; \textit{R1 - R3}: $p < .001$; \textit{R1 - R4}: $p < .001$; \textit{R2 - R3}: $p < .001$; \textit{R2 - R4}: $p < .001$; \textit{R3 - R4}: $p < .001$). For content \textit{Soldier}, significant differences were found, for codec \textit{C1}, for all bitrate touples, bar \textit{R3 - R4} ($p = 0.648$; for all other touples, $p < .001$); conversely, for codec \textit{C2}, the only bitrates who did not exhibit significant differences were \textit{R1 - R2} ($p = 0.373$; for all other touples, $p < .001$). This seems to indicate that generally, for codec \textit{C1} statistical differences in score distributions are usually found between lower bitrates, whereas the highest bitrates have a more uniform behaviour; however, for codec \textit{C2}, more differences can be found across score distribution for all the bitrates, bar certain cases (such as \textit{Loot} and \textit{Soldier}) for which the lowest bitrates are deemed equivalent. 
Several other combinations of the three factors under exam were deemed statistically significant, more than what we could cover in this pages: we refer interested readers to the appendix for a full coverage of the post-hoc interaction results.
\color{black}

\subsection{Additional questionnaires and interaction data}
\subsubsection{IPQ \& SSQ Questionnaires}
For T1 and T2, the collected IPQ data under each subscale are all normally distributed as examined by the Shapiro-Wilk test ($p>0.05$). A paired sample t-test was applied to check the differences between 3DoF and 6DoF in terms of SP, INV, REAL and G. For T1, there was a significant difference in SP between 3DoF (M=4.13, SD=0.92) and 6DoF (M=5.04, SD=0.67), t(26)=-4.44, $p<.001$, Cohen's d = 0.52 and also a significant difference in G between 3DoF (M=4.11, SD=1.28) and 6DoF (M=4.96, SD=1.13), t(26)=-2.60, $p<.01$, Cohen's d = 0.64. For T2, SP was also significantly different in 3DoF (M=4.16, SD=1.17) and 6DoF (M=4.83, SD=1.12), t(24)=-3.48, $p<.01$, Cohen's d = 0.45 and so was G between 3DoF (M=4.20, SD=1.61) and 6DoF (M=5.08, SD=1.19), t(24)=-3.56, $p<.01$, Cohen's d = 0.71. Other factors showed no significant differences between 3DoF and 6DoF in both T1 and T2.

With respect to SSQ, no significant differences ($p>0.05$) were found between 3DoF and 6DoF in terms of cybersickness. We further tested whether there were order effects in experiencing cybersickness, where half of the participants started with 6DoF as the first condition and 3DoF as the second, and the remainder the inverse. No significant differences ($p>0.05$) were found for any order effects in experiencing cybersickness.

\begin{figure}
\begin{center}
    \hf
    \subfloat[T1]{\includegraphics[width=0.9\wtc]{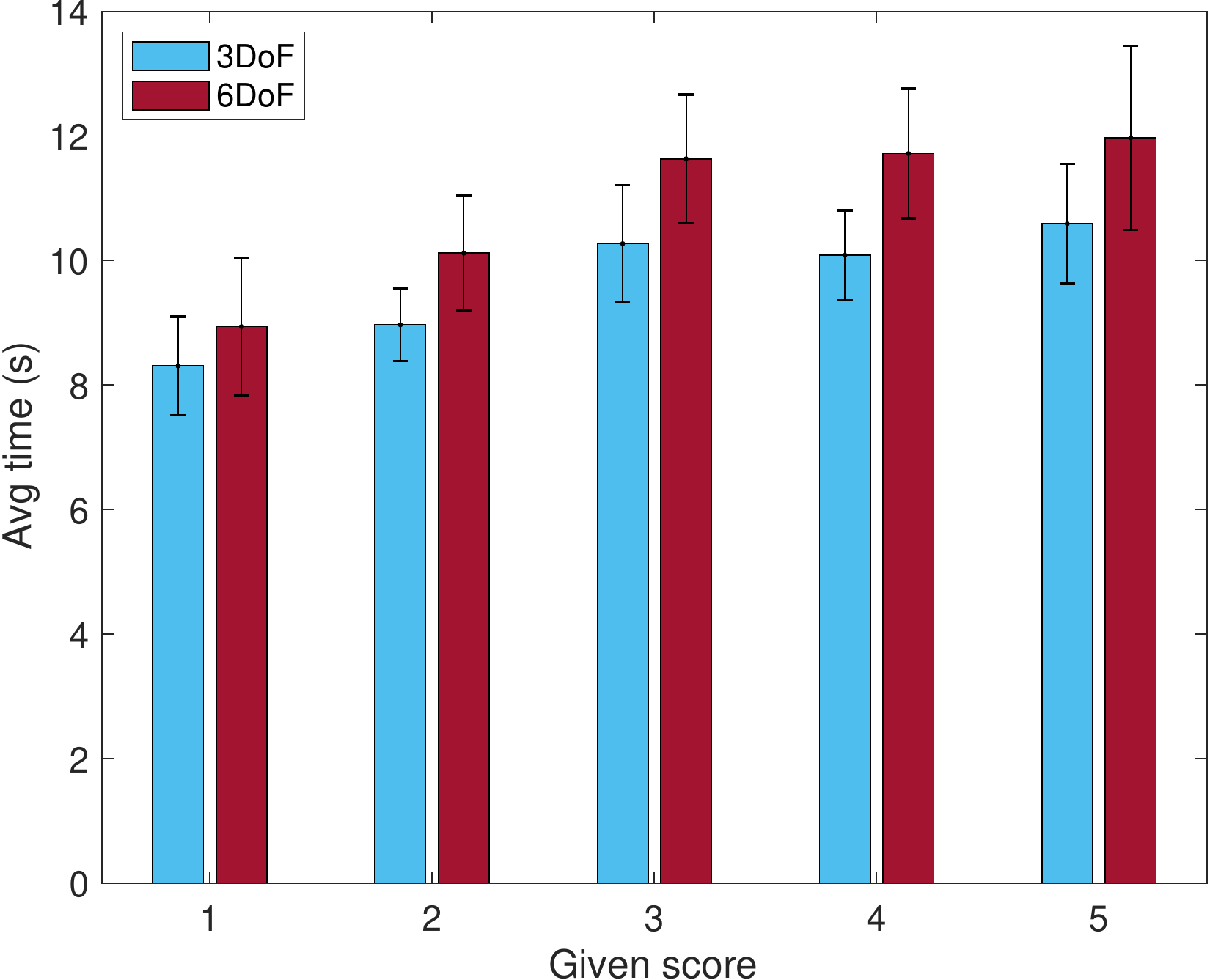}}\hf
    \subfloat[T2]{\includegraphics[width=0.9\wtc]{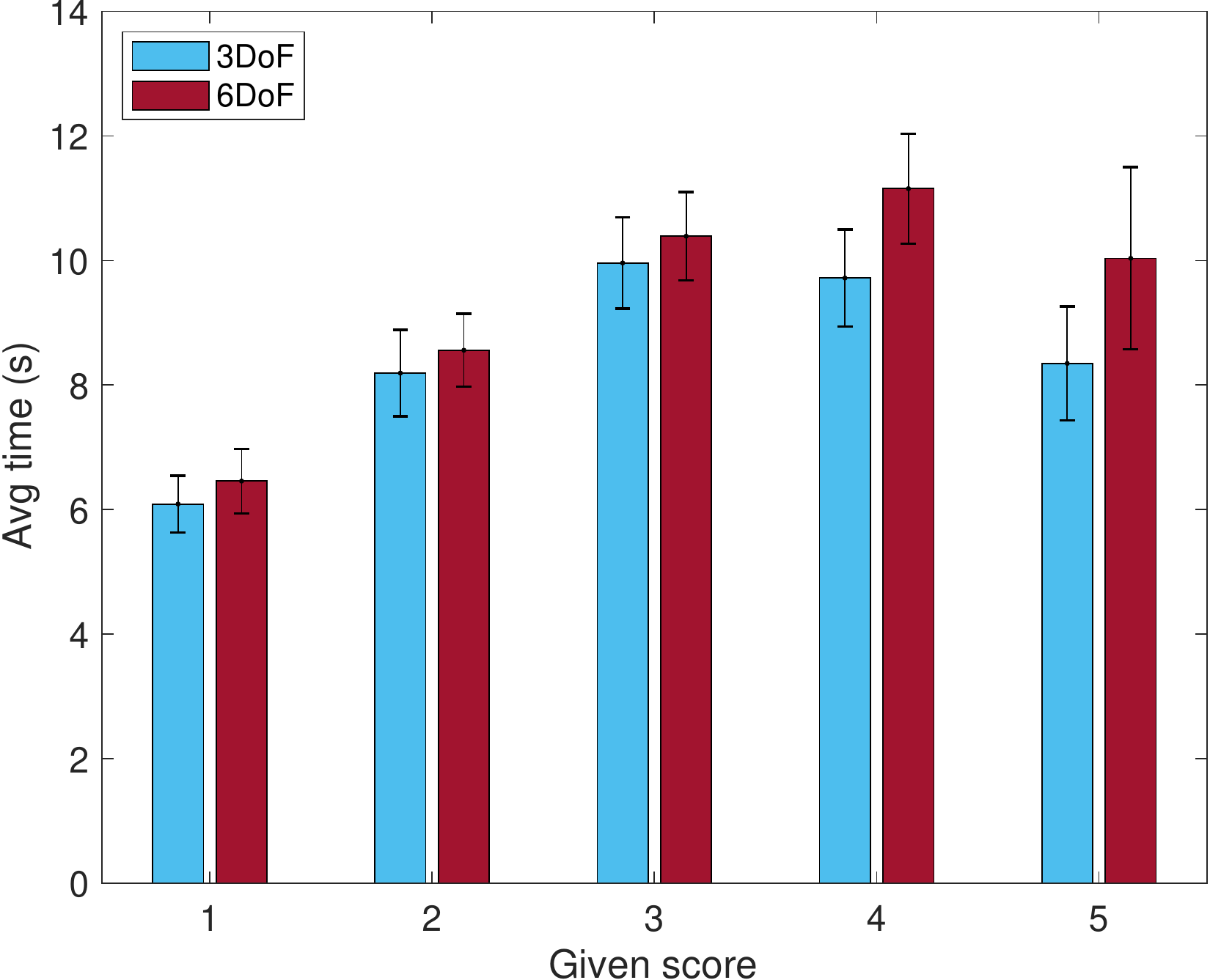}}\hf
    \caption{Average time spent looking at the sequence (in seconds) and relative CIs, against score given to the sequence, for 3DoF (blue) and 6DoF (red), in test T1 (left) and T2 (right).}
    \label{fig:avgtime}
    \end{center}
    \vspace{-0.7cm}
\end{figure}

\subsubsection{Interaction time}
Interaction time was found to be strongly correlated with MOS values in a study conducted on light field image quality assessment~\cite{viola2017new}, as well as in studies conducted with point cloud contents in interactive environments~\cite{alexiou2020pointxr}. In particular, it was found that users tended to spend more time interacting with contents at high quality, whereas for low quality scores, less time was spent looking at the contents.  In order to see whether similar trends could be observed in our data, we compared the average time spent watching the sequence in 3DoF and 6DoF, separately for each quality score given by the participants. Since no interactivity was present in the 2DTV condition, the data was discarded for the analysis. Results are shown in Fig.~\ref{fig:avgtime}. A positive trend can be observed between the given score and the average time spent looking at the sequence, with the exception of score 5, which for test T2 shows a negative trend with respect to the time. However, it should be considered that on average, a small percentage of scores equal to 5 were given in test T2 (10\% of the total scores), thus, variations may be due to the difference in sample size.
It is also worth noting that, on average, participants spent more time looking at the sequences in 6DoF, with respect to the 3DoF case. Indeed, several participants pointed out that the lowest scores were the fastest to be given, whereas for higher quality, it was harder to decide on the rating.

\subsubsection{Interviews}
The interviews were only conducted for the VR conditions, due to time limitations. We asked the same interview questions for T1 and T2; so, we combined the interview transcripts of 52 participants (T1=27, T2=25). \textcolor{black}{Participants were labelled as T1P1-T1P27 or T2P1-T2P25}. The categorized answers are presented as follows:

\textbf{\textit{Factors considered when assessing quality.}} 57\% of the participants mentioned that they assessed the quality based on three criteria: 1) overall outline and pattern distortion on body and on clothes, 2) natural gestures and movements of the digital humans, and 3) visual artifacts such as blockiness, blurriness, and extraneous floating artifacts. \textcolor{black}{As T2P3 commented, \textit{``I paid attention to the blurriness of the clothing patterns, the fingers, and whether the gestures or movements were smooth. ''} } 48\% of the participants mentioned the quality assessment criteria are content related, who agreed that it is easier to spot artifacts for the content with complex patterns (e.g., \textit{Long dress}) and dominant colors (e.g., \textit{Red and black}) than the content with uniformed colors (e.g., \textit{Soldier} and \textit{Sarge}). \textcolor{black}{T2P5 said, \textit{``The two ladies were easier [to assess the visual quality], because their clothes have strong colors. The man playing keyboard or something is quite difficult. His clothes were mainly monotone.''}} 46\% of the participants considered facial expressions as an unignorable factor for quality assessment, which they believe is an important cue for social connectedness. \textcolor{black}{For example, T1P15 mentioned, \textit{``The robotic lady was static and had no [facial] expression at all. It was difficult to tell the difference [between different quality levels]. '' } Similarly, T2P8 said, \textit{``If I could see her [the lady with red dress] smile and her teeth, I gave the score of fair to good. ''}} For the extraneous floating artifacts (e.g., bubbles flickering outside the digital humans), 23\% found it very annoying and lowered the overall quality for the content, but a few participants (8\%) thought these artifacts do not influence their quality judgement. \textcolor{black}{T1P4 commented, \textit{``The flickering blocks were annoying and distracting at the beginning, but later I got used to them. It felt like watching an old movie.''}}

\textbf{\textit{Difficulties in assessment.} }42\% of the participants pointed out the difficulties in assessing the quality, especially for the high quality contents, which are not perfect and still have missing details like blurry faces or wrong fingers. 15\% of the participants specifically pointed out that it is difficult to distinguish between quality level 3 to 5. \textcolor{black}{As T2P21 mentioned, \textit{``It was difficult to give [score] 5, the best ones were still missing many details like fingers or feet. I was hesitating all the time between [score] 4 and 5.''}} 17\% of the participants commented that it gradually became easier in rating the quality when they adapted to the contents. So, the second viewing condition was easier for them. \textcolor{black}{For example, T1P12 said, \textit{``I noticed that the rating got easier as I got familiar to the quality levels, and had seen the best and the worst qualities.''}}

\textbf{\textit{Comparison between 3DoF and 6DoF.}} 52\% of the participants preferred 6DoF, because it allowed them to move closer to examine the details (e.g., shoes and fingers). They felt more realistic when walking in the virtual space. However, they also commented that 3DoF offered a fixed distance between them and digital humans, enabling a more stable and focused assessment. \textcolor{black}{For example, T2P13 commented, \textit{``Walking around [6DoF] allowed me to get really close [to the digital humans] and see more details, like pixel sizes, shoes, fingers. It felt more realistic. Sitting down [3DoF] has a distance [between me and the digital humans]. Sometimes, I found it difficult to assess the quality.''}} 21\% of the participants preferred relaxation and passiveness in 3DoF, because they did not find much differences between 3DoF and 6DoF in terms of quality assessment, but they found 3DoF is less nauseous than 6DoF. \textcolor{black}{As T1P13 mentioned, \textit{``I felt more focused, secured and relaxed when sitting down. I got worried to be trapped by the cables. I also felt more dizzy when I walked around [in VR].''}}

\section{Discussion}
\label{sec:discussion}

\subsection{Testing in VR}
\textcolor{black}{Results of our experiment indicate a very small, if not negligible, effect of viewing condition on the distribution of the scores. In particular, the viewing condition was deemed to have a significant effect on the distribution of the scores for the first set of contents we tested, revealing significant differences between VR testing and the 2DTV counterpart, with small effect sizes; for the second set of contents, no significant difference was spotted. 
The implications of it are twofold. On one hand, results show that the score distributions follow similar trends in VR with respect to passive video consumption, thus confirming the validity of testing volumetric contents on traditional 2D screens, as it has been done for the majority of tests performed in the literature. More specifically, no significant interaction was found, for both tests, between viewing condition and codec under exam, indicating that differences among compression solutions are not affected by the choice of displaying device and interaction paradigm. However, the interaction between viewing condition and contents was found to be significant for test T1: in particular, results indicate that differences among contents vary depending on the visualization medium. This might be due to a variety of factors: the possibility of interacting with the content, moving closer to inspect details; the presence of a fixed viewing point, which allows for easier comparison; the absence of confounding factors such as simulator-induced sickness or novelty effect. Particular care should be adopted in choosing contents depending on the type of test that needs to be carried, making sure that artifacts are visible at the distance that is selected for passive viewing, for example.}

\textcolor{black}{The second implication involves the constituents of quality of experience. The MOS of visual quality is only one of the factors that influences the quality of experience of a given user; other factors, such as presence and immersion, are equally important in determining the enjoyment of a given user using the system. Even though small or no effect of viewing condition on the MOS distribution was found, results of the IPQ revealed a strong effect of viewing condition on spatial presence and general sense of being there. Such factors should be taken into consideration when designing new experiments: if visual quality is the main constituent that needs to be assessed, traditional screens might be substituted (with caution) for VR assessment; however, if other factors need to be evaluated, such choice might have a larger impact.}

\subsection{Datasets}
Despite the rich literature in point cloud acquisition and compression, few point cloud datasets are publicly available. This is especially true when considering point cloud datasets depicting photo-realistic humans. One of the most popular and widely used full-body dataset, created by 8i Labs~\cite{8i:dataset}, consists of only 4 individual contents, whereas the HHI Fraunhofer dataset has 1 individual content~\cite{hhi:dataset}.  
In the context of point cloud compression, such scarcity of available data may lead to compression solutions being designed, optimized and tested while considering a considerably narrow range of input data, thus leading to algorithms that are overfitted to the specifics of the acquisition method used to obtain the contents.
The consequences of such a scenario are reflected in our results. Whereas for the contents assessed in test T2 a large difference was observed between codec V-PCC and the MPEG anchor, for the contents in test T1 the gap was markedly lower, and indeed the significance of the effect of the codec selection had a smaller effect size for test T1 with respect to test T2, as seen in section~\ref{subsec:qoe}. Test T2 consisted of contents that had been used in multiple quality assessment experiments~\cite{torlig2018novel, alexiou2019towards, Alexiou2019comprehensive, da2019point, perry2020quality}, notably including the performance evaluation of the upcoming MPEG standard~\cite{schwarz2018emerging}. On the other hand, test T1 included contents that have not been used so far in assessment of point cloud compression solutions. The discrepancies in the results of the subjective quality assessment campaign indicate that performance gains may vary considerably when new contents are evaluated.
A larger body of contents depicting digital humans, involving several acquisition technologies, is needed in order to properly design, train and evaluate new compression solutions in a robust way.

\subsection{Personal preferences and bias}
Subjective evaluation experiments are complicated by many aspects of human psychology and viewing conditions, such as participants' vision ability, translation of quality perception into ranking scores, adaptations and personal preferences for contents. Through carefully following the ITU-T Recommendations P.913~\cite{ITUTP913}, we are able to control some of the aspects. For example, eliminate the scores given by the participants with vision problems; train participants to help them understand the quality levels; randomize the stimuli and viewing conditions to minimize the order effects. However, we noticed that personal preferences towards certain contents are difficult to control. Satgunam et al. \cite{satgunam2013factors}) found that their participants were divided into two preference groups: prefer sharper content versus smoother content. Similarly, Kortum and Sullivan\cite{kortum2010effect} found that the "desirability" of participants had an impact on video quality responses, with a more desirable video clip being given a higher rating. In our experiments, content \textit{Queen} is generally given lower ratings with respect to the other contents. In the interviews, 40\% of the participants showed their preference towards \textit{Soldier}, due to his high-resolution facial features, unitoned clothes and natural movements, whereas 27\% expressed dislike towards \textit{Queen}, because of her lifeless look and static gestures. \textcolor{black}{Body motion has been shown to be important in increasing the naturalness of virtual characters, especially when involving complex motions~\cite{neff2016hand}. Moreover, several studies have highlighted the importance of matching the appearance of the avatar to the naturalness of the motion, indicating that “appropriateness” between visual presentation and gesture plays an important role in realism~\cite{kucherenko2021large, ferstl2021human}. Our findings suggests that realism and naturalness of interaction might have an impact on the perceived visual quality of the contents as well.}
Quality assessment may need to be adjusted based on content and viewer preferences, and offering training with different contents, \textcolor{black}{as well as account for factors such as realism, naturalness of motion, and uncanny valley effect.}

\subsection{Technological constraints and limitations}
The two codecs used in this experiment introduce different distortions during compression. As the MPEG anchor codec uses the octree data structure to represent geometry, the number of points in the decoded cloud varies exponentially based on the tree depth. 
Thus, at lower bitrates, the decoded point clouds are quite sparse, and when the point size is increased to make them appear watertight, they have a block-y appearance. 
However, the low delay encoding and decoding of this codec makes it suitable for real time applications such as social VR.
%
%
On the other hand, the V-PCC codec leverages existing 2D video codecs to compress both geometry and color, which introduces noise in terms of extraneous objects, and general geometric artifacts such as misaligned seams (see Figure~\ref{fig:pcccreation}).
However, the approach yields better results at low bitrates, as demonstrated in our results. The codec is optimized for human perception of 2D video, which might lead to undesired results when applied to 3D objects in VR. The mapping from 3D to 2D is critical to codec performance, which explains why the encoding phase has high complexity. Decoding has a lower delay, as it benefits from hardware acceleration of video decoders on GPUs, making this approach suitable for on-demand streaming.

One of the main shortcoming of both compression solutions lays in their inability to reach visually-lossless quality, as demonstrated by our results. Achieving a visually pleasant result is of paramount importance for the market adoption of the technology; indeed, poor visual quality might lead consumers to tune off from the experience altogether~\cite{conviva}. Visual perception should be taken into account when designing compression solutions, especially at high bitrates, to ensure that in absence of strict bandwidth constraints, excellent quality can be achieved.





\subsection{Rendering environment considerations} 

\textcolor{black}{Previous research on subjective assessment of dynamic point cloud contents has been primarily conducted in desktop setups~\cite{mpeg:cfp, zerman2019subjective}, whereas VR/AR technologies have been employed with static content~\cite{alexiou2017towards, alexiou2019towards}, or limited amount of dynamic contents~\cite{zerman2021user}.
However, placing dynamic contents to be rendered in a VR/AR environment in real time, in order for users to interact, adds several technological constraints. In previous research on subjective assessment of point cloud content such as the MPEG standardisation activity \cite{mpeg:cfp}, participants were asked to view a video of the point cloud sequence rendered from a predefined camera path. The same approach was used by van der Hooft et al. to subjectively and objectively assess the quality of adaptive streaming for point cloud contents~\cite{van2020objective}; however, the influence of camera paths on objective quality assessment was shown to be significant in~\cite{subramanyam2020user}. 
In order to allow users to interact with the content in a VR/AR environment, the dynamic point clouds need to be rendered in real time. This is, however, extremely resource intensive, and poses many technical constraints. 
For our test in particular, the point clouds needed to be stored as uncompressed files, to avoid confounding factors with the compression solutions under evaluation.}
%
%
\textcolor{black}{The point cloud files were stored as binary PLY files to allow faster read from disk. 
Yet, real-time progressive loading of the sequences was not possible, as the loading operation was interfering with the rendering, thus leading to drops in the frame rate. Waiting for each sequence to be loaded, on the other hand, was unattainable,
as the read time from disk would mean long waiting times between one sequence and another, thus adding to subjects' fatigue.
To fix the issue, we decided to load all sequences in physical memory before the test. However, this impacted the amount of sequences that could be tested in one session, as well as the length of each. 
New systems for rendering point clouds in real time, while respecting the constraints introduced by quality assessment scenarios, should be developed in order to foster research in the field.}

\subsection{Protocols for subjective assessment in VR}
Choosing the right methodology to follow in order to collect users' opinions is a delicate matter, as it can influence the statistical power of the collected score, and in some cases lead to difference in results. Single stimulus methodologies, in particular, lead to larger CIs with respect to double stimulus methodologies, and are more subject to be influenced by individual content preference~\cite{ITUTP913}. An early study comparing single and double stimulus methodologies for the evaluation of colorless point cloud contents indicated that the latter was more consistent in recognizing the level of impairment, as relative differences facilitate the rating task~\cite{alexiou2017performance}. However, the study pointed out that the single stimulus methodology shows more discrimination power for compression-like artifacts, albeit at the cost of wider CIs.

Double stimulus methodologies, while commonly used in video quality assessment and widely adopted in 2D-based quality assessment of point cloud contents~\cite{da2019point, schwarz2018emerging, Alexiou2019comprehensive}, are tricky to adopt in VR technology, due to the difficulties in displaying both contents simultaneously in a perceptually satisfying way~\cite{perrin2017measuring}, while ensuring a fair comparison between the contents under evaluation. When dealing with interactive methodologies, in particular, synchronous display of any modification in viewport is usually enforced, to ensure that the two contents are always visible at the same condition~\cite{viola2017new, Alexiou2019comprehensive}. This is clearly challenging to implement in a 6DoF scenario, in which users are free to change their position in the VR space at any given time. Positioning the two contents side by side in the same virtual space would mean that, at any given time, they are seen from two different angles; the same problem would arise when temporal sequencing is employed. A toggle-based method like the one proposed in~\cite{perrin2017measuring} is not applicable to moving sequences, as different frames would be seen between stimuli.

In our study, we saw that content preference had an impact on the ratings, as several contents were deemed of lower quality, as the scores given to the HR exemplify. Such bias resulted in a reduced rating range for the contents. Results of the interviews also pointed out that naturalness of gestures were an important criteria in assessing the visual quality. Such components would not be normally evaluated in a double stimulus scenario; however, they are important in understanding how human perception reacts to digital humans. 


\section{Conclusion}
\label{sec:conclusions}

In this paper, we extend our previous work by comparing the performance of the point cloud compression standard V-PCC against an octree-based anchor codec (MPEG anchor). Participants were invited to assess the quality of digital humans represented as dynamic point clouds, 
in 2DTV screen, as well as VR with 3DoF and 6DoF conditions. Results indicate a small effect of viewing condition on the final scores for one of the two sets of contents under test. Moreover, results show that codec V-PCC has a more favorable performance than the MPEG anchor, especially at low bit-rates. For the highest bit-rate, the two codecs are often statistically equivalent. 
The contents under test appear to have a significant influence on how the scores are distributed; thus, new data sets are needed in order to comprehensively evaluate compression distortions. Moreover, current encoding solutions, while efficient at low bitrates, are unable to provide visually lossless results, even when large volumes of data are available, revealing significant shortcomings in point cloud compression.   
We also point out that commonly-used double stimulus methodologies for quality evaluation often reduce the rating task to a difference recognition, while insights on the quality of the original contents are missed. \textcolor{black}{The raw data is made available at the following link: \url{https://github.com/cwi-dis/2DTV\_VR\_QoE}}.

\bibliographystyle{spmpsci}      

\bibliography{reference.bib}

\begin{thebibliography}{10}
\providecommand{\url}[1]{{#1}}
\providecommand{\urlprefix}{URL }
\expandafter\ifx\csname urlstyle\endcsname\relax
  \providecommand{\doi}[1]{DOI~\discretionary{}{}{}#1}\else
  \providecommand{\doi}{DOI~\discretionary{}{}{}\begingroup
  \urlstyle{rm}\Url}\fi

\bibitem{conviva}
{OTT: Beyond Entertainment Consumer Survey Report}.
\newblock \url{https://www.conviva.com/research/ott-beyond-entertainment/}

\bibitem{mpeg:cfp}
Call for proposals for point cloud compression iso/iec jtc1/sc29 wg11 n16732,
  geneva ch  (2017)

\bibitem{alexiou2017performance}
Alexiou, E., Ebrahimi, T.: On the performance of metrics to predict quality in
  point cloud representations.
\newblock In: Applications of Digital Image Processing XL, vol. 10396, p.
  103961H. International Society for Optics and Photonics (2017)

\bibitem{alexiou2018impact}
Alexiou, E., Ebrahimi, T.: Impact of visualisation strategy for subjective
  quality assessment of point clouds.
\newblock In: 2018 IEEE International Conference on Multimedia \& Expo
  Workshops (ICMEW), pp. 1--6. IEEE (2018)

\bibitem{alexiou2020towards}
Alexiou, E., Tung, K., Ebrahimi, T.: Towards neural network approaches for
  point cloud compression.
\newblock In: Applications of Digital Image Processing XLIII, vol. 11510, p.
  1151008. International Society for Optics and Photonics (2020)

\bibitem{alexiou2017towards}
Alexiou, E., Upenik, E., Ebrahimi, T.: Towards subjective quality assessment of
  point cloud imaging in augmented reality.
\newblock In: 2017 IEEE 19th International Workshop on Multimedia Signal
  Processing (MMSP), pp. 1--6. IEEE (2017)

\bibitem{Alexiou2019comprehensive}
Alexiou, E., Viola, I., Borges, T.M., Fonseca, T.A., de~Queiroz, R.L.,
  Ebrahimi, T.: A comprehensive study of the rate-distortion performance in
  mpeg point cloud compression.
\newblock APSIPA Transactions on Signal and Information Processing \textbf{8},
  27 (2019).
\newblock \doi{10.1017/ATSIP.2019.20}.
\newblock \urlprefix\url{http://infoscience.epfl.ch/record/272124}

\bibitem{alexiou2019towards}
Alexiou, E., Xu, P., Ebrahimi, T.: Towards modelling of visual saliency in
  point clouds for immersive applications.
\newblock In: 26th IEEE International Conference on Image Processing (ICIP)
  (2019)

\bibitem{alexiou2020pointxr}
Alexiou, E., Yang, N., Ebrahimi, T.: Pointxr: A toolbox for visualization and
  subjective evaluation of point clouds in virtual reality.
\newblock In: 2020 Twelfth International Conference on Quality of Multimedia
  Experience (QoMEX), pp. 1--6. IEEE (2020)

\bibitem{cao2021compression}
Cao, C., Preda, M., Zakharchenko, V., Jang, E.S., Zaharia, T.: Compression of
  sparse and dense dynamic point clouds--methods and standards.
\newblock Proceedings of the IEEE  (2021)

\bibitem{cohen2016point}
Cohen, R.A., Tian, D., Vetro, A.: Point cloud attribute compression using 3-d
  intra prediction and shape-adaptive transforms.
\newblock In: 2016 Data Compression Conference (DCC), pp. 141--150. IEEE (2016)

\bibitem{8i:dataset}
d'Eon, E., Harrison, B., Myers, T., Chou, P.A.: {8i Voxelized Full Bodies - A
  Voxelized Point Cloud Dataset, ISO/IEC JTC1/SC29 Joint WG11/WG1 (MPEG/JPEG)
  input document WG11M40059/WG1M74006, Geneva}  (2017)

\bibitem{hhi:dataset}
Ebner, T., Feldmann, I., Schreer, O., Kauff, P., v.~Unger, T.: {HHI Point cloud
  dataset of a boxing trainer, ISO/IEC JTC1/SC29 Joint WG11/WG1 (MPEG/JPEG)
  input document MPEG2018/m42921, Ljubljana}  (2018)

\bibitem{ebrahimi2016jpeg}
Ebrahimi, T., Foessel, S., Pereira, F., Schelkens, P.: Jpeg pleno: Toward an
  efficient representation of visual reality.
\newblock Ieee Multimedia \textbf{23}(4), 14--20 (2016)

\bibitem{elkin2021aligned}
Elkin, L.A., Kay, M., Higgins, J.J., Wobbrock, J.O.: An aligned rank transform
  procedure for multifactor contrast tests.
\newblock arXiv preprint arXiv:2102.11824  (2021)

\bibitem{ferstl2021human}
Ferstl, Y., Thomas, S., Guiard, C., Ennis, C., McDonnell, R.: Human or robot?
  investigating voice, appearance and gesture motion realism of conversational
  social agents.
\newblock In: Proceedings of the 21st ACM International Conference on
  Intelligent Virtual Agents, pp. 76--83 (2021)

\bibitem{guarda2019deep}
Guarda, A.F., Rodrigues, N.M., Pereira, F.: Deep learning-based point cloud
  coding: A behavior and performance study.
\newblock In: 2019 8th European Workshop on Visual Information Processing
  (EUVIP), pp. 34--39. IEEE (2019)

\bibitem{guarda2019point}
Guarda, A.F., Rodrigues, N.M., Pereira, F.: Point cloud coding: Adopting a deep
  learning-based approach.
\newblock In: 2019 Picture Coding Symposium (PCS), pp. 1--5. IEEE (2019)

\bibitem{guarda2020deep}
Guarda, A.F., Rodrigues, N.M., Pereira, F.: Deep learning-based point cloud
  geometry coding: Rd control through implicit and explicit quantization.
\newblock In: 2020 IEEE International Conference on Multimedia \& Expo
  Workshops (ICMEW), pp. 1--6. IEEE (2020)

\bibitem{van2020objective}
van~der Hooft, J., Vega, M.T., Timmerer, C., Begen, A.C., De~Turck, F., Schatz,
  R.: Objective and subjective qoe evaluation for adaptive point cloud
  streaming.
\newblock In: 2020 twelfth international conference on quality of multimedia
  experience (QoMEX), pp. 1--6. IEEE (2020)

\bibitem{MPEG-GPCC-standard_ISO}
{ISO/IEC DIS 23090-5}: {Geometry-based point cloud compression}.
\newblock International Organization for Standardization

\bibitem{MPEG-VPCC-standard_ISO}
{ISO/IEC DIS 23090-5}: {Visual volumetric video-based coding (V3C) and
  video-based point cloud compression (V-PCC)}.
\newblock International Organization for Standardization (2021)

\bibitem{ITURBT500}
{ITU-R BT.500-13}: Methodology for the subjective assessment of the quality of
  television pictures.
\newblock International Telecommunication Union (2012)

\bibitem{ITUTP910}
{ITU-T P.910}: {Subjective video quality assessment methods for multimedia
  applications}.
\newblock International Telecommunication Union (2008)

\bibitem{ITUTP913}
{ITU-T P.913}: {Methods for the subjective assessment of video quality, audio
  quality and audiovisual quality of Internet video and distribution quality
  television in any environment}.
\newblock International Telecommunication Union (2016)

\bibitem{jackins1980oct}
Jackins, C.L., Tanimoto, S.L.: Oct-trees and their use in representing
  three-dimensional objects.
\newblock Computer Graphics and Image Processing \textbf{14}(3), 249--270
  (1980)

\bibitem{jang2019video}
Jang, E.S., Preda, M., Mammou, K., Tourapis, A.M., Kim, J., Graziosi, D.B.,
  Rhyu, S., Budagavi, M.: Video-based point-cloud-compression standard in mpeg:
  from evidence collection to committee draft [standards in a nutshell].
\newblock IEEE Signal Processing Magazine \textbf{36}(3), 118--123 (2019)

\bibitem{matthew_kay_2019_2556415}
Kay, M., Wobbrock, J.: mjskay/artool: Artool 0.10.6 (2019).
\newblock \doi{10.5281/zenodo.2556415}.
\newblock \urlprefix\url{https://doi.org/10.5281/zenodo.2556415}

\bibitem{kennedy1993simulator}
Kennedy, R.S., Lane, N.E., Berbaum, K.S., Lilienthal, M.G.: Simulator sickness
  questionnaire: An enhanced method for quantifying simulator sickness.
\newblock The international journal of aviation psychology \textbf{3}(3),
  203--220 (1993)

\bibitem{kortum2010effect}
Kortum, P., Sullivan, M.: The effect of content desirability on subjective
  video quality ratings.
\newblock Human factors \textbf{52}(1), 105--118 (2010)

\bibitem{kucherenko2021large}
Kucherenko, T., Jonell, P., Yoon, Y., Wolfert, P., Henter, G.E.: A large,
  crowdsourced evaluation of gesture generation systems on common data: The
  genea challenge 2020.
\newblock In: 26th International Conference on Intelligent User Interfaces, pp.
  11--21 (2021)

\bibitem{mammou2017pcc}
Mammou, K.: {PCC test model category 2 v0}.
\newblock ISO/IEC JTC1/SC29/ WG11 N17248 \textbf{1} (2017)

\bibitem{meagher1982geometric}
Meagher, D.: Geometric modeling using octree encoding.
\newblock Computer graphics and image processing \textbf{19}(2), 129--147
  (1982)

\bibitem{mekuria2017design}
Mekuria, R., Blom, K., Cesar, P.: Design, implementation, and evaluation of a
  point cloud codec for tele-immersive video.
\newblock IEEE Transactions on Circuits and Systems for Video Technology
  \textbf{27}(4), 828--842 (2017)

\bibitem{neff2016hand}
Neff, M.: Hand gesture synthesis for conversational characters.
\newblock Handbook of Human Motion pp. 1--12 (2016)

\bibitem{perrin2017measuring}
Perrin, A.F., Bist, C., Cozot, R., Ebrahimi, T.: Measuring quality of
  omnidirectional high dynamic range content.
\newblock In: Applications of Digital Image Processing XL, vol. 10396, p.
  1039613. International Society for Optics and Photonics (2017)

\bibitem{perry2020quality}
Perry, S., Cong, H.P., da~Silva~Cruz, L.A., Prazeres, J., Pereira, M.,
  Pinheiro, A., Dumic, E., Alexiou, E., Ebrahimi, T.: Quality evaluation of
  static point clouds encoded using mpeg codecs.
\newblock In: 2020 IEEE International Conference on Image Processing (ICIP),
  pp. 3428--3432. IEEE (2020)

\bibitem{quach2019learning}
Quach, M., Valenzise, G., Dufaux, F.: Learning convolutional transforms for
  lossy point cloud geometry compression.
\newblock In: 2019 IEEE International Conference on Image Processing (ICIP),
  pp. 4320--4324. IEEE (2019)

\bibitem{quach2020improved}
Quach, M., Valenzise, G., Dufaux, F.: Improved deep point cloud geometry
  compression.
\newblock In: 2020 IEEE 22nd International Workshop on Multimedia Signal
  Processing (MMSP), pp. 1--6. IEEE (2020)

\bibitem{queiroz:raht}
Queiroz, R.D., Chou, P.A.: {Compression of 3D Point Clouds Using a
  Region-Adaptive Hierarchical Transform}.
\newblock IEEE Transactions on Image Processing 25  (2016)

\bibitem{satgunam2013factors}
Satgunam, P.N., Woods, R.L., Bronstad, P.M., Peli, E.: Factors affecting
  enhanced video quality preferences.
\newblock IEEE Transactions on Image Processing \textbf{22}(12), 5146--5157
  (2013)

\bibitem{schubert2003sense}
Schubert, T.W.: The sense of presence in virtual environments: A
  three-component scale measuring spatial presence, involvement, and realness.
\newblock Zeitschrift f{\"u}r Medienpsychologie \textbf{15}(2), 69--71 (2003)

\bibitem{schwarz2018emerging}
Schwarz, S., Preda, M., Baroncini, V., Budagavi, M., Cesar, P., Chou, P.A.,
  Cohen, R.A., Krivoku{\'c}a, M., Lasserre, S., Li, Z., et~al.: {Emerging MPEG
  standards for point cloud compression}.
\newblock IEEE Journal on Emerging and Selected Topics in Circuits and Systems
  \textbf{9}(1), 133--148 (2018)

\bibitem{da2019point}
da~Silva~Cruz, L.A., Dumi{\'c}, E., Alexiou, E., Prazeres, J., Duarte, R.,
  Pereira, M., Pinheiro, A., Ebrahimi, T.: Point cloud quality evaluation:
  Towards a definition for test conditions.
\newblock In: 2019 Eleventh International Conference on Quality of Multimedia
  Experience (QoMEX), pp. 1--6. IEEE (2019)

\bibitem{subramanyam2020comparing}
Subramanyam, S., Li, J., Viola, I., Cesar, P.: Comparing the quality of highly
  realistic digital humans in 3dof and 6dof: A volumetric video case study.
\newblock In: 2020 IEEE Conference on Virtual Reality and 3D User Interfaces
  (VR), pp. 127--136. IEEE (2020)

\bibitem{subramanyam2020user}
Subramanyam, S., Viola, I., Hanjalic, A., Cesar, P.: User centered adaptive
  streaming of dynamic point clouds with low complexity tiling.
\newblock In: Proceedings of the 28th ACM International Conference on
  Multimedia, pp. 3669--3677 (2020)

\bibitem{torlig2018novel}
Torlig, E.M., Alexiou, E., Fonseca, T.A., de~Queiroz, R.L., Ebrahimi, T.: A
  novel methodology for quality assessment of voxelized point clouds.
\newblock In: Applications of Digital Image Processing XLI, vol. 10752, p.
  107520I. International Society for Optics and Photonics (2018)

\bibitem{tran2019subjective}
TT~Tran, H., Ngoc, N.P., Pham, C.T., Jung, Y.J., Thang, T.C.: A subjective
  study on user perception aspects in virtual reality.
\newblock Applied Sciences \textbf{9}(16), 3384 (2019)

\bibitem{viola2017new}
Viola, I., Ebrahimi, T.: A new framework for interactive quality assessment
  with application to light field coding.
\newblock In: Applications of Digital Image Processing XL, vol. 10396, p.
  103961F. International Society for Optics and Photonics (2017)

\bibitem{wang2021lossy}
Wang, J., Zhu, H., Liu, H., Ma, Z.: Lossy point cloud geometry compression via
  end-to-end learning.
\newblock IEEE Transactions on Circuits and Systems for Video Technology
  (2021)

\bibitem{wobbrock2011aligned}
Wobbrock, J.O., Findlater, L., Gergle, D., Higgins, J.J.: {The Aligned Rank
  Transform for nonparametric factorial analyses using only ANOVA procedures}.
\newblock In: Proceedings of the SIGCHI conference on human factors in
  computing systems, pp. 143--146. ACM (2011)

\bibitem{zerman2019subjective}
Zerman, E., Gao, P., Ozcinar, C., Smolic, A.: Subjective and objective quality
  assessment for volumetric video compression.
\newblock In: Fast track article for IST International Symposium on Electronic
  Imaging 2019: Image Quality and System Performance XVI proceedings (2019)

\bibitem{zerman2021user}
Zerman, E., Kulkarni, R., Smolic, A.: User behaviour analysis of volumetric
  video in augmented reality.
\newblock In: 2021 13th International Conference on Quality of Multimedia
  Experience (QoMEX), pp. 129--132. IEEE (2021)

\bibitem{zerman2020textured}
Zerman, E., Ozcinar, C., Gao, P., Smolic, A.: Textured mesh vs coloured point
  cloud: A subjective study for volumetric video compression.
\newblock In: 2020 Twelfth International Conference on Quality of Multimedia
  Experience (QoMEX), pp. 1--6. IEEE (2020)

\bibitem{zhang:gft}
Zhang, C., Florencio, D., Loop, C.: Point cloud attribute compression with
  graph transform.
\newblock Image Processing (ICIP), 2014 IEEE International Conference on
  (2014)

\bibitem{zhang2014subjective}
Zhang, J., Huang, W., Zhu, X., Hwang, J.N.: {A subjective quality evaluation
  for 3D point cloud models}.
\newblock In: 2014 International Conference on Audio, Language and Image
  Processing, pp. 827--831. IEEE (2014)

\end{thebibliography}

\newpage
\onecolumn
\section*{Appendix A}
\label{sec:appendix}


\end{longtable}



%

\end{document}